\newif\ifapj
\apjtrue

\ifapj
  \documentclass[iop]{emulateapj_update}
\else
  \documentclass[12pt,preprint]{aastex}
\fi

\usepackage{amsmath,apjfonts,amsfonts,amssymb,fullpage,enumerate,graphicx,
  longtable,float,times,textcomp,verbatim,url,nicefrac}
\def\apjl{ApJ}%
\def\apjs{ApJS}%
\def\ao{Appl.~Opt.}%
\def\aap{A\&A}%
\ifapj
\fi

\tabletypesize{\footnotesize}

\newcommand{\drivelc}{1367\,\AA}
\newcommand{\citePei}{Paper V}
\newcommand{\HST}{{\it HST}}

\newcommand{\Vlag}{$2.04 \pm 0.21$}
\newcommand{\alphafit}{$0.79 \pm 0.22$}
\newcommand{\betafit}{$0.99 \pm 0.14$}

\shorttitle{AGN STORM III. Optical Photometric Observations}
\shortauthors{Fausnaugh et al.}
\begin{document}

\title{Space Telescope and Optical Reverberation Mapping Project.\\
III.\ Optical Continuum Emission and Broad-Band Time Delays in NGC 5548}

\author{M.~M.~Fausnaugh\altaffilmark{1},
K.~D.~Denney\altaffilmark{1,2,3},
A.~J.~Barth\altaffilmark{4},
M.~C.~Bentz\altaffilmark{5},
M.~C.~Bottorff\altaffilmark{6},
M.~T.~Carini\altaffilmark{7},
K.~V.~Croxall\altaffilmark{1,2},
G.~De~Rosa\altaffilmark{1,2,8},
M.~R.~Goad\altaffilmark{9},
Keith~Horne\altaffilmark{10},
M.~D.~Joner\altaffilmark{11},
S.~Kaspi\altaffilmark{12,13},
M.~Kim\altaffilmark{14},
S.~A.~Klimanov\altaffilmark{15}, 
C.~S.~Kochanek\altaffilmark{1,2},
D.~C.~Leonard\altaffilmark{16},
H.~Netzer\altaffilmark{12},
B.~M.~Peterson\altaffilmark{1,2},
K.~Schn\"{u}lle\altaffilmark{17},
S.~G.~Sergeev\altaffilmark{18},
M.~Vestergaard\altaffilmark{19,20},
W.-K.~Zheng\altaffilmark{21},  
Y.~Zu\altaffilmark{1,22},
M.~D.~Anderson\altaffilmark{5},
P.~Ar\'{e}valo\altaffilmark{23},
C.~Bazhaw\altaffilmark{5},
G.~A.~Borman\altaffilmark{18},
T.~A.~Boroson\altaffilmark{24},
W.~N.~Brandt\altaffilmark{25,26,27},
A.~A.~Breeveld\altaffilmark{28},
B.~J.~Brewer\altaffilmark{29},
E.~M.~Cackett\altaffilmark{30},
D.~M.~Crenshaw\altaffilmark{5},
E.~Dalla~Bont\`{a}\altaffilmark{31,32},
A.~De~Lorenzo-C\'{a}ceres\altaffilmark{10},
M.~Dietrich\altaffilmark{33,34},
R.~Edelson\altaffilmark{35},
N.~V.~Efimova\altaffilmark{15},
J.~Ely\altaffilmark{8},
P.~A.~Evans\altaffilmark{9},
A.~V.~Filippenko\altaffilmark{21} ,
K.~Flatland\altaffilmark{16},
N.~Gehrels\altaffilmark{36},
S.~Geier\altaffilmark{37,38,39},
J.~M.~Gelbord\altaffilmark{40,41},
L.~Gonzalez\altaffilmark{16},
V.~Gorjian\altaffilmark{42},
C.~J.~Grier,\altaffilmark{1,25,26},
D.~Grupe\altaffilmark{43},
P.~B.~Hall\altaffilmark{44},
S.~Hicks\altaffilmark{7}, 
D.~Horenstein\altaffilmark{5},
T.~Hutchison\altaffilmark{6},
M.~Im\altaffilmark{45},
J.~J.~Jensen\altaffilmark{19},
J.~Jones\altaffilmark{5},
J.~Kaastra\altaffilmark{46,47,48},
B.~C.~Kelly\altaffilmark{49},
J.~A.~Kennea\altaffilmark{24},
S.~C.~Kim\altaffilmark{14},
K.~T.~Korista\altaffilmark{50},
G.~A.~Kriss\altaffilmark{8,51},
J.~C.~Lee\altaffilmark{14},
P.~Lira\altaffilmark{52},
F.~MacInnis\altaffilmark{6},
E.~R.~Manne-Nicholas\altaffilmark{5},
S.~Mathur\altaffilmark{1,2},
I.~M.~M$^{\rm c}$Hardy\altaffilmark{53},
C.~Montouri\altaffilmark{54},
R.~Musso\altaffilmark{6},
S.~V.~Nazarov\altaffilmark{18},
R.~P.~Norris\altaffilmark{5},
J.~A.~Nousek\altaffilmark{25},
D.~N.~Okhmat\altaffilmark{18},
A.~Pancoast\altaffilmark{55,56},
I.~Papadakis\altaffilmark{57,58},
J.~R.~Parks\altaffilmark{5},
L.~Pei\altaffilmark{4},
R.~W.~Pogge\altaffilmark{1,2},
J.-U.~Pott\altaffilmark{17},
S.~E.~Rafter\altaffilmark{13,59},
H.-W.~Rix\altaffilmark{17},
D.~A.~Saylor\altaffilmark{5},
J.~S.~Schimoia\altaffilmark{1,60},
M.~Siegel\altaffilmark{24},
M.~Spencer\altaffilmark{11},
D.~Starkey\altaffilmark{10},
H.-I.~Sung\altaffilmark{14},
K.~G.~Teems\altaffilmark{5},
T.~Treu\altaffilmark{49,61,62},
C.~S.~Turner\altaffilmark{5},
P.~Uttley\altaffilmark{63},
C.~Villforth\altaffilmark{64},
Y.~Weiss\altaffilmark{13}
J.-H.~Woo\altaffilmark{45},
H.~Yan\altaffilmark{65},
and S.~Young\altaffilmark{35}
}

\altaffiltext{1}{Department of Astronomy, The Ohio State University,
  140 W 18th Ave, Columbus, OH 43210, USA}
 \altaffiltext{2}{Center for Cosmology and AstroParticle Physics, The
Ohio State University, 191 West Woodruff Ave, Columbus, OH 43210, USA}
\altaffiltext{4}{Department of Physics and Astronomy, 4129 Frederick
Reines Hall, University of California, Irvine, CA 92697, USA}
\altaffiltext{5}{Department of Physics and Astronomy, Georgia State
University, 25 Park Place, Suite 605, Atlanta, GA 30303, USA}
\altaffiltext{6}{Fountainwood Observatory, Department of Physics FJS 149,
Southwestern University, 1011 E. University Ave., Georgetown, TX 78626, USA}
\altaffiltext{7}{Department of Physics and Astronomy, Western Kentucky University,
1906 College Heights Blvd \#11077, Bowling Green, KY 42101, USA}
\altaffiltext{8}{Space Telescope Science Institute, 3700 San Martin
Drive, Baltimore, MD 21218, USA}
\altaffiltext{9}{University of Leicester, Department of Physics and Astronomy,
Leicester, LE1 7RH, UK}
\altaffiltext{10}{SUPA Physics and Astronomy, University of
St. Andrews, Fife, KY16 9SS Scotland, UK}
\altaffiltext{11}{Department of Physics and Astronomy, N283 ESC, Brigham Young University,
Provo, UT 84602-4360, USA}
\altaffiltext{12}{School of Physics and Astronomy, Raymond and Beverly Sackler Faculty of Exact
Sciences, Tel Aviv University, Tel Aviv 69978, Israel}
\altaffiltext{13}{Physics Department, Technion, Haifa 32000, Israel}
\altaffiltext{14}{Korea Astronomy and Space Science Institute, Republic of Korea}
\altaffiltext{15}{Pulkovo Observatory, 196140 St.\ Petersburg, Russia}
\altaffiltext{16}{Department of Astronomy, San Diego State University, San Diego, CA
  92182-1221, USA}
\altaffiltext{17}{Max Planck Institut f\"{u}r Astronomie, K\"{o}nigstuhl 17,
D--69117 Heidelberg, Germany} 
\altaffiltext{18}{Crimean Astrophysical Observatory, P/O Nauchny,
Crimea 298409, Russia}
\altaffiltext{19}{Dark Cosmology Centre, Niels Bohr Institute,
University of Copenhagen, Juliane Maries Vej 30, DK-2100 Copenhagen,
Denmark}
\altaffiltext{20}{Steward Observatory, University of Arizona, 933
North Cherry Avenue, Tucson, AZ 85721, USA}
\altaffiltext{21}{Department of Astronomy, University of California, Berkeley, CA 94720-3411, USA}
\altaffiltext{22}{Department of Physics, Carnegie Mellon University,
5000 Forbes Avenue, Pittsburgh, PA 15213, USA}
\altaffiltext{23}{Instituto de F\'{\i}sica y Astronom\'{\i}a, Facultad
de Ciencias, Universidad de Valpara\'{\i}so, Gran Bretana N 1111,
Playa Ancha, Valpara\'{\i}so, Chile}
\altaffiltext{24}{Las Cumbres Global Telescope Network, 6740 Cortona Drive, Suite 102,
Santa Barbara, CA  93117, USA}
\altaffiltext{25}{Department of Astronomy and Astrophysics, Eberly
College of Science, The Pennsylvania State University, 525 Davey Laboratory,
University Park, PA 16802, USA}
\altaffiltext{26}{Institute for Gravitation and the Cosmos, The
Pennsylvania State University, University Park, PA 16802, USA}
\altaffiltext{27}{Department of Physics, The Pennsylvania State
University, 104 Davey Lab, University Park, PA 16802}
\altaffiltext{28}{Mullard Space Science Laboratory, University College
London, Holmbury St. Mary, Dorking, Surrey RH5 6NT, UK}
\altaffiltext{29}{Department of Statistics, The University of
Auckland, Private Bag 92019, Auckland 1142, New Zealand}
\altaffiltext{30}{Department of Physics and Astronomy, Wayne State University,
666 W. Hancock St, Detroit, MI 48201, USA}
\altaffiltext{31}{Dipartimento di Fisica e Astronomia ``G. Galilei,''
Universit\`{a} di Padova, Vicolo dell'Osservatorio 3, I-35122 Padova,
Italy}
\altaffiltext{32}{INAF-Osservatorio Astronomico di Padova, Vicolo
dell'Osservatorio 5 I-35122, Padova, Italy}
\altaffiltext{33}{Department of Physics and Astronomy, Ohio
University, Athens, OH 45701, USA}
\altaffiltext{34}{Department of Earth, Environment, and Physics, Worcester
State University, 486 Chandler Street, Worcester, MA 01602, USA}
\altaffiltext{35}{Department of Astronomy, University of Maryland,
College Park, MD 20742-2421, USA}
\altaffiltext{36}{Astrophysics Science Division, NASA Goddard Space
Flight Center, Greenbelt, MD 20771, USA}
\altaffiltext{37}{Instituto de Astrof\'{\i}sica de Canarias, 38200 La Laguna,
Tenerife, Spain}
\altaffiltext{38}{Departamento de Astrof\'{\i}sica, Universidad de La Laguna,
E-38206 La Laguna, Tenerife, Spain}
\altaffiltext{39}{Gran Telescopio Canarias (GRANTECAN), 38205 San Crist\'{o}bal de La
Laguna, Tenerife, Spain}
\altaffiltext{40}{Spectral Sciences Inc., 4 Fourth Ave., Burlington,
MA 01803, USA}
\altaffiltext{41}{Eureka Scientific Inc., 2452 Delmer St. Suite 100,
Oakland, CA 94602, USA}
\altaffiltext{42}{MS 169-327, Jet Propulsion Laboratory, California Institute of Technology, 
4800 Oak Grove Drive, Pasadena, CA 91109, USA}
\altaffiltext{43}{Space Science Center, Morehead State University, 235
Martindale Dr., Morehead, KY 40351, USA}
\altaffiltext{44}{Department of Physics and Astronomy, York
University, Toronto, ON M3J 1P3, Canada }
\altaffiltext{45}{Astronomy Program, Department of Physics \& Astronomy, 
Seoul National University, Seoul, Republic of Korea}
\altaffiltext{46}{SRON Netherlands Institute for Space Research,
Sorbonnelaan 2, 3584 CA Utrecht, The Netherlands}
\altaffiltext{47}{Department of Physics and Astronomy, Univeristeit
Utrecht, P.O. Box 80000, 3508 Utrecht, The Netherlands}
\altaffiltext{48}{Leiden Observatory, Leiden University, PO Box 9513,
2300 RA Leiden, The Netherlands}
\altaffiltext{49}{Department of Physics, University of California,
Santa Barbara, CA 93106, USA}
\altaffiltext{50}{Department of Physics, Western Michigan University,
1120 Everett Tower, Kalamazoo, MI 49008-5252, USA}
\altaffiltext{51}{Department of Physics and Astronomy, The Johns
Hopkins University, Baltimore, MD 21218, USA}
\altaffiltext{52}{Departamento de Astronomia, Universidad de Chile,
Camino del Observatorio 1515, Santiago, Chile}
\altaffiltext{53}{University of Southampton, Highfield, Southampton,
SO17 1BJ, UK}
\altaffiltext{54}{DiSAT, Universita dell'Insubria, via Valleggio 11, 22100, Como, Italy}
\altaffiltext{55}{Harvard-Smithsonian Center for Astrophysics, 60 Garden Street, Cambridge, MA 02138, USA}
\altaffiltext{57}{Department of Physics and Institute of Theoretical
and Computational Physics, University of Crete, GR-71003 Heraklion,
Greece}
\altaffiltext{58}{IESL, Foundation for Research and Technology,
GR-71110 Heraklion, Greece}
\altaffiltext{59}{Department of Physics, Faculty of Natural Sciences, University of Haifa,
Haifa 31905, Israel}
\altaffiltext{60}{Instituto de F\'{\i}sica, Universidade Federal do
Rio do Sul, Campus do Vale, Porto Alegre, Brazil}
\altaffiltext{61}{Department of Physics and Astronomy, University of
California, Los Angeles, CA 90095-1547, USA}
\altaffiltext{63}{Astronomical Institute `Anton Pannekoek,' University
of Amsterdam, Postbus 94249, NL-1090 GE Amsterdam, The Netherlands}
\altaffiltext{64}{University of Bath, Department of Physics, Claverton Down, BA2 7AY, Bath, United Kingdom}
\altaffiltext{65}{Department of Physics and Astronomy, University of
Missouri, Columbia, MO 65211, USA}

\footnotetext[3]{NSF Postdoctoral Research Fellow}
\footnotetext[56]{Einstein Fellow}
\footnotetext[62]{Packard Fellow}

\begin{abstract}
  We present ground-based optical photometric monitoring data for NGC
  5548, part of an extended multi-wavelength reverberation mapping
  campaign.  The light curves have nearly daily cadence from 2014
  January to July in nine filters (\emph{BVRI} and \emph{ugriz}).
  Combined with ultraviolet data from the \emph{Hubble Space Telescope} and
  \emph{Swift}, we confirm significant time delays between the
  continuum bands as a function of wavelength, extending the
  wavelength coverage from 1158\,\AA\ to the $z$ band ($\sim\!
  9160$\,\AA).  We find that the lags at wavelengths longer than the
  {\it V} band are equal to or greater than the lags of 
  high-ionization-state emission lines (such as He\,{\sc ii}\,$\lambda
  1640$ and $\lambda 4686$), suggesting that the continuum-emitting
  source is of a physical size comparable to the inner broad-line
  region (BLR).  The trend of lag with wavelength is broadly consistent with
  the prediction for continuum reprocessing by an accretion disk with
  $\tau \propto \lambda^{4/3}$.  However, the lags also imply a disk
  radius that is 3 times larger than the prediction from standard
  thin-disk theory, assuming that the bolometric luminosity is 10\% of
  the Eddington luminosity ($L = 0.1L_{\rm Edd}$).  Using optical
  spectra from the Large Binocular Telescope, we estimate the bias of
  the interband continuum lags due to BLR emission
  observed in the filters.  We find that the bias for filters with
  high levels of BLR contamination ($\sim\!  20\%$) can be important
  for the shortest continuum lags, and likely has a significant impact
  on the {\it u} and {\it U} bands owing to Balmer continuum
  emission.
\end{abstract}
\section{Introduction}
The continuum emission of radio-quiet active galactic nuclei (AGN) is
believed to originate in an accretion disk around a supermassive
black hole (SMBH).  At accretion rates and masses appropriate for
SMBHs, geometrically thin, optically thick accretion disks have
maximum temperatures of $\sim\!  10^5$--$10^6$ K, naturally
accounting for the characteristic peak of AGN spectral energy
distributions (SEDs) in the far ultraviolet
\citep[UV;][]{Burbidge1967,Shakura1973,Sheilds1978}.  However, a large
variety of competing models of the accretion flow exist, such as
thick-disk geometries at extremely super- or sub-Eddington accretion
rates \citep{Abramowicz1988,Narayan1995}.  In addition, AGN exhibit
nonthermal X-ray emission, which requires a hot plasma component or
``corona'' \citep[e.g., ][]{Haardt1991,Chakrabarti1995}.  The
potential configurations and complex interplay between the hot corona
and accretion disk admit a wide range of models with many free
parameters, and searching for the unique observational signatures of a
given disk model is very challenging \citep[and references
therein]{Sun1989, Laor1997, Karatkar1999, Vestergaard2001, Telfer2002,
  Kishimoto2004}.

Reverberation mapping \citep[RM;
][]{Blandford1982,Peterson1993,Peterson2014} can provide insight into
the structure of the accretion disk, and has become a standard tool
for AGN astrophysics over the last 25 years \citep[and references
therein]{Clavel1991,Peterson1991,Horne1991, Kaspi2000, Peterson2004,
  Bentz2009b, Denney2010, Grier2013, Pancoast2014,Pei2014, Barth2015}.
The basic principle of RM is that emission at two different
wavelengths is causally connected, so that the time delay (or lag)
between two light curves represents the light-crossing time within the
system, and thereby provides a straightforward measurement of the
system's physical size.  For example, because the AGN continuum powers
the prominent emission lines observed in Seyfert galaxy/quasar
spectra, the time delays between continuum and broad-line light curves
are commonly used to determine the physical extent of the
line-emitting gas (the so-called broad-line region, BLR).

In a similar way, RM techniques can be used to constrain the physical
processes governing AGN continuum emission.  X-ray emission from the
corona may irradiate and heat the accretion disk.  If the corona is
relatively compact and centrally located, the UV and optical emission
would be expected to respond to the incident X-ray flux, ``echoing''
the X-ray light curve after a time delay corresponding to the 
light-travel time across the disk \citep{Krolik1991}.  On the other hand,
X-ray light curves would be expected to lag behind UV and optical
light curves if the X-rays are produced by Comptonization of thermal
UV/optical disk photons \citep{Haardt1991}.  Observational
investigations of the relation between X-ray and UV/optical emission
have produced ambiguous results.  X-rays have been found to lead the
optical emission by one to several days in some objects
\citep[e.g.,][]{Arevalo2009, Breedt2010, Shappee2014, Troyer2015}, but
the X-ray variability on long ($>1$ year) timescales cannot always
account for the optical variations \citep{Uttley2003, Breedt2009}. In
addition, other studies find no long-term X-ray/optical correlations
\citep{Maoz2002}, or find optical variations that lead the X-rays on
shorter timescales \citep[$\sim\!  15$ days, ][]{Marshall2008}.

RM can also reveal information about the size and structure of the
continuum-emitting source.  Emission from different portions of the
disk peaks at different wavelengths depending on the local disk
temperature.  By translating the wavelength of the continuum emission
into a characteristic temperature, time delays between continuum light
curves can be used to map the temperature profile of the disk.  The
first statistically significant interband time delays were found in
NGC 7469 by \citet{Wanders1997} and \citet{Collier1998}.
\cite{Sergeev2005} carried out intensive optical monitoring of 14 AGN
and found evidence that longer wavelengths lag shorter-wavelength
emission.  More recent continuum RM campaigns have used the
\emph{Swift} observatory \citep{Gehrels2004} to obtain unprecedentedly
well-sampled light curves across X-ray, near-UV, and optical
wavelengths: \citet{Shappee2014} observed NGC 2617 with {\it Swift} on
a nearly daily basis for several months in 2014, while
\citet{McHardy2014} monitored NGC 5548 with $\sim\! 2$ day cadence for
approximately 2 years (excepting seasonal gaps).  These studies found
trends of lag with wavelength that are well fit by the expectation for
X-ray/far-UV reprocessing.

The present study is the third in a series describing the results of
the AGN Space Telescope and Optical Reverberation Mapping (STORM)
project, an intensive, multi-wavelength monitoring campaign of NGC
5548.  The AGN STORM campaign is anchored by daily far-UV observations
using the Cosmic Origins Spectrograph (COS; \citealt{Green2012}) on
the \emph{Hubble Space Telescope} (\emph{HST}).  \citet[hereinafter
Paper I]{DeRosa2015} give a complete introduction to the project and
an analysis of the \emph{HST} data.  The COS program was complemented
by a four-month broad-band photometric monitoring campaign using
\emph{Swift}, the first results of which are presented by
\citet[hereinafter Paper II]{Edelson2015}.  The \emph{Swift} campaign
achieved $\sim\!0.5$ day cadence and detected significant lags between
the UV and optical continua, which follow the expected lag-wavelength
relation of a thin accretion disk ($\tau \propto \lambda^{4/3}$).

Supplementing these space-based observations are ground-based optical
monitoring programs.  The present study details the optical broad-band
photometric monitoring component, which extends the analysis in Paper
II using data in nine optical filters with $\lesssim\! 1$ day cadence
for seven months.  The similarly intensive ground-based spectroscopic
monitoring will be presented by Pei et al. (in prep., hereinafter Paper
V).  In terms of cadence, temporal baseline, and wavelength coverage,
the combination of UV and optical observations of the AGN STORM
project represents the most complete RM experiment ever conducted.

The present work has three primary goals.  The first is to directly
compare the far-UV and optical light curves of NGC 5548 over a
concurrent monitoring period.  The far-UV light curve
($\sim$1350\,\AA) is expected to closely trace the true ionizing
continuum ($\leq$912\,\AA), while the optical continuum
($\sim$5100\,\AA) appears to be delayed and somewhat smoothed compared
to the UV emission.  Since ground-based RM campaigns use the optical
continuum as a proxy for the driving continuum light curve,
understanding how the continuum emission changes as a function of
wavelength is important for understanding any systematic effects in
optical RM experiments.  The second goal is to search for time delays
between the UV and optical data, in an attempt to probe the structure
of the continuum-emitting region. However, because broad-band filters
pick up spectral features that arise in the BLR (e.g., strong emission
lines), and these features have large lags relative to the underlying
continuum (several days for a Seyfert galaxy such as NGC 5548),
interband lags estimated from broad-band photometry may be biased
indicators of the accretion-disk size.  Therefore, our final goal is
to estimate the impact of BLR emission on the observed interband time
delays.

In \S2, we describe the observations, data reduction, flux
calibration, and general properties of the ground-based photometric
light curves.  In \S3, we describe our time-series analysis, measuring
the lag as a function of wavelength of the broad-band filters.  We
then explore the impact of BLR emission on the observed interband
lags in \S4.  Finally, in \S5, we discuss our results, and we
summarize our conclusions in \S6.  Where relevant, we assume a
standard cosmological model with $\Omega_{m} = 0.28$,
$\Omega_{\Lambda} = 0.72$, and $H_{0} = 70$ km\,s$^{-1}$
\citep{Komatsu2011}.

\section{Observations and Data Reduction}
In conjunction with the \emph{HST COS} UV RM campaign described in
Paper I, NGC 5548 was observed between 2013 December and 2014 August
by 16 ground-based observatories in optical broad-band filters:
Johnson/Cousins \emph{BVRI} and Sloan Digital Sky Survey (SDSS) 
\emph{ugriz}.  A short
description of each telescope, the relevant imager, and the number of
contributed epochs is given in Table \ref{tab:telescopes}.  All
observatories followed a common reduction protocol: images were first
overscan-corrected, bias-subtracted, and flat-fielded following
standard procedures.  The reduced data, as well as nightly calibration
frames, raw images, and observing logs, were then uploaded to a
central repository, and the image quality was assessed by eye.  Images
taken in reasonable atmospheric conditions and free of obvious
reduction errors were analyzed as described below.
\ifapj
  \begin{deluxetable*}{llrlcccrr}
\else
  \begin{deluxetable}{llrlcccrr}
  \rotate
\fi

  \tablewidth{0pt} \tablecaption{Contributing
    Observatories \label{tab:telescopes}} 
  \tablehead{
    \colhead{Observatory Name}& \colhead{Obs ID}& \colhead{Aperture}&
    \colhead{Detector}& \colhead{Pixel Scale}& \colhead{Field of View}& 
    \colhead{Observing Period}&\colhead{Filters}&\colhead{Epochs }
  } \startdata
  Bohyunsan Optical Astronomy                       & BOAO & 1.8m&e2v CCD231-84  &0\farcs21 &$15' \times 15'$  &March-April &\emph{V} &5 \\
  \phantom{aaa}Observatory\\
  Crimean Astrophysical      & CrAO & 0.7m&AP7p CCD       &1\farcs76 &$15'\times 15'$   &Dec-June    &\emph{BVRI} &76 \\
  \phantom{aaa}Observatory \\  
  Fountainwood Observatory                & FWO  & 0.4m&SBIG 8300M     &0\farcs35 &$19'\times 15'$   &Jan-August  &\emph{V} &60 \\
  Hard Labor Creek Observatory            & HLCO & 0.5m&Apogee USB/Net &0\farcs75 &$25'\times 25'$   &April-June  &\emph{V} &27 \\
  La Silla Observatory                    &GROND & 2.2m&Gamma-ray Burst Optical/&0\farcs33&$5'\times 5'$&Feb-July &\emph{griz} &6\\
                                          &      &     & Near-IR Detector\\
  Las Cumbres Observatory           &LCOGT & 1.0m&SBIGSTX-16803/ &0\farcs23 &$16' \times 16'$ &Jan-August  &\emph{BV\,ugriz} &263 \\
  \phantom{aaa}Global Telescope Network  &  & &Sinistro CCD-486BI & 0\farcs39& $27' \times 27'$  \\
  Lick Observatory Katzman        &KAIT & 0.8m& AP7 CCD        &0\farcs80&$7' \times 7'$     &Jan-July    &\emph{V}&80 \\
  \phantom{aaa}Automatic Imaging Telescope\\
  Liverpool Telescope                     &LT   & 2.0m& e2v CCD 231    &0\farcs15&$10'\times 10'$    &Feb-August  &\emph{ugriz}&120 \\
  Maidanak Observatory                    &MO15 & 1.5m& SNUCAM         &0\farcs27&$18'\times 18'$    &April-August&\emph{BVR}&45 \\
  Mt. Laguna Observatory                   &MLO  & 1.0m&CCD2005         &0\farcs41&$14'\times 14'$    &June-August &\emph{V}&10 \\
  Mt. Lemmon Optical  &LOAO & 1.0m&KAF-4301E      &0\farcs68 &$22' \times 22'$   &Feb-July    &\emph{V}&26 \\
  \phantom{aaa}Astronomy Observatory\\
  Nordic Optical Telescope                &NOT  & 2.5m&e2V CCD42-40   &0\farcs19  &$6' \times  6'$   &April       &\emph{V}&3 \\
  Robotically Controlled         & RCT  & 1.3m&SITe CCD       &0\farcs30 &$10' \times 10'$  &Dec-May     &\emph{BV}&55 \\
  \phantom{aaa}Telescope\\
  Svetloe Observatory                     &SvO  & 0.4m& ST-7XME CCD    &2\farcs00&$12' \times 8'$   &Jan-May     &\emph{BVRI} &49 \\
  West Mountain Observatory               & WMO  & 0.9m&Finger Lakes PL-3041-UV &0\farcs61&$21'\times 21'$    &Jan-July    &\emph{BVR} &44 \\
  Wise Observatory                        & WC18 & 0.5m&STL-6303E CCD  &1\farcs47&$75'\times 50'$    &Dec-July    &\emph{BVRI} &126 \\
\enddata

\ifapj
  \end{deluxetable*}
\else
  \end{deluxetable}
\fi

\subsection{Differential Photometry}
The analysis is based on the {\tt ISIS} image-subtraction software
package \citep{Alard1998}.  Images are first registered to a common
coordinate system, and the images with the lowest backgrounds and best
seeing are combined into a high signal-to-noise ratio (S/N) ``reference'' 
image. The other images are then rescaled to match the effective exposure
time of the reference image.  Next, the reference image is convolved
with a spatially variable kernel to match the point-spread function (PSF) of each individual
epoch, and then subtracted to leave the variable flux in each pixel.
We use {\tt ISIS}'s built-in photometry package to extract light
curves from the subtracted images at the position of the AGN in NGC
5548, in units of differential counts relative to the reference image.
Because each telescope/filter/detector combination has slightly
different properties (pixel scales, fields of view, gains, etc.), we
built reference frames and subtracted images for each unique dataset.
This procedure corrects for variable seeing conditions and removes
nonvariable sources such as host-galaxy starlight, allowing a clean
measurement of the variable AGN flux.

\subsection{Measurement Uncertainties}
The formal errors found by {\tt ISIS} sometimes underestimate the full
uncertainties because they only account for local Poisson error
contributions.  In order to estimate more reliable measurement
uncertainties, we examined the residual fluxes of stars in the
subtracted images, and rescaled the formal {\tt ISIS} errors to be
consistent with the observed scatter.  Our method is similar to that
of \citet[\S4.1]{Hartman2004}.

We first used {\tt ISIS} to extract differential light curves at the
positions of each unsaturated star in the reference images.  For stars
with constant flux in time, the distribution of residual fluxes at
each epoch serves as an estimate of the uncertainty in the
subtraction. Since we are only concerned with the magnitude of the
residuals, we first take their absolute value.  We then divide these
values by their formal {\tt ISIS} uncertainties, so that the resulting
ratios indicate the factor by which the true uncertainties are
underestimated.  We set a minimum value of 1.0 for this ratio, since
the uncertainty cannot be smaller than the local photon noise.
Finally, we multiply the formal uncertainty for the AGN at the
matching epoch by the median of the rescaling factors from all stars.
The procedure ensures that the measurement uncertainty in a given
image is consistent with the observed scatter of the subtracted stars.
The median rescaling factor for all images was 2.9, while 75\% of the
rescaling factors are less than 6.6 and 98\% are less than 25.0.  The
remaining 2\% have rescaling factors between 30 and 87.  The poorest
subtractions result when {\tt ISIS} cannot accurately construct the
image PSF, usually because the image has too few stars.

To assess the effectiveness of this method, we adjusted the stars'
uncertainties by the derived rescaling factor for each image, and then
checked the goodness-of-fit for a constant-flux model of each star.
The goodness-of-fit is calculated by
\begin{align}\label{chi2/dof}
\chi^2/{\rm dof} = \frac{1}{N-1}\sum_i^N\left(\frac{c_i - \bar c}{\sigma_i}\right)^2,
\end{align}
where ${\rm dof} = N - 1$ is the number of degrees of freedom of the
fit, $c_i$ is the counts in the light curve at epoch $i$, $\sigma_i$
is the rescaled uncertainty, and $\bar c$ is the mean counts of the
light curve.  90\% of the rescaled values of $\chi^2/{\rm dof}$ are
between 0.32 and 2.09, and the distribution peaks at $0.81$, somewhat
lower than would be expected for purely Gaussian statistics.
This may indicate that our rescaling method is slightly overestimating
the measurement uncertainties.  However, given our large dataset, we
can afford to be conservative in this regard.

Data from the Katzman Automatic Imaging Telescope
\citep[KAIT;][]{filippenko2001} and the {\it u}-band data from the 
Liverpool Telescope
\citep[LT;][]{Steele2004} required a different treatment since these
images have 10 stars or fewer, which are not enough to provide robust
estimates of the error-rescaling factors. We instead calculated global
rescaling factors from all available epochs, rather than individual
corrections from single images.  Using Equation \ref{chi2/dof}, we
calculate $\chi^2/{\rm dof}$ for all available stars, using the
unscaled {\tt ISIS} uncertainties for $\sigma_i$.  We then multiplied
the uncertainties of the AGN light curve by the average value of
$(\chi^2/{\rm dof})^{1/2}$.  We found that the mean rescaling factor
was 8.99 for the KAIT data and 2.23 for the {\it u}-band LT data.
Although this method does not account for epochs with high-quality
subtractions, we find the cautious approach preferable to
underestimating the uncertainties.

\subsection{Intercalibration of Light Curves}
In order to combine differential light curves in the same filter but
from different telescopes, it is necessary to intercalibrate the light
curves to a common flux scale.  This accounts for the different mean
flux levels and analog-to-digital unit (ADU) definitions between the
reference images, as well as small differences in filter transmission
functions, detector efficiencies, etc., of the many telescopes.  We
model the difference of any two light curves by a multiplicative
rescaling factor $a$ and an additive shift $b$.  While it is trivial
to solve for these parameters by matching epochs where the fluxes are
known to be equal, no two observations occur at precisely the same
time and it is therefore necessary to interpolate the light curves.
Furthermore, this method can only treat two light curves at a time,
and therefore loses information by ignoring the global probability of
the ensemble calibration parameters for all telescopes.  In order to
address both of these problems, we built a full statistical model of
the intercalibrated light curve using the software package {\tt
  JAVELIN}, following the {\tt SPEAR} formalism of \citet{Zu2011}.

{\tt JAVELIN} models the light curves as a damped random walk (DRW).
Although recent studies using \emph{Kepler} light curves have shown
that the DRW overpredicts the amplitude of AGN continuum variability
on short timescales \citep{Edelson2014, Kasliwal2015}, the DRW
provides an adequate description of the observed light curves for the
noise properties and cadence/timescales of this study (we
quantitatively verify this claim in the Appendix, but see also
\citealt{Kelly2009,MacLeod2010,Zu2013}).  In brief, points sampled
from a DRW have an exponential covariance matrix, which is described
by an amplitude $\sigma_{\rm DRW}$ that characterizes the strength of
short-term variations, and a damping timescale $\tau_{\rm DRW}$ over
which the light curve becomes decoherent.  Using a Markov Chain Monte
Carlo (MCMC) calculation, we simultaneously fit for the shifts and
rescaling factors of all light curves in a single filter.  We also fit
for $\sigma_{\rm DRW}$, but our light curves do not have a
sufficiently long temporal baseline to meaningfully constrain
$\tau_{\rm DRW}$.  We therefore fixed $\tau_{\rm DRW}$ = 164 days, so
as to match the value determined from multiyear historical light
curves of NGC 5548 \citep{Zu2011}.  The model provides a well-defined
and self-consistent means of interpolating all the light curves
simultaneously (see \citealt{Zu2011,Li2013} for further details).

Our fitting procedure requires one
light curve to be chosen to define the flux scale and mean flux level
of the resulting intercalibration, so that this reference light curve
is assigned a shift of $0$ and a rescaling factor of $1$.  In the
Johnson \emph{BVRI} bands, we use the Wise C18
\citep[WC18;][]{Brosch2008} data as the calibration light curve, owing
to its dense temporal sampling, long baseline, and large number of
comparison stars ($>$400).  For the SDSS \emph{ugriz} bands, we use
the LT light curves, since they have the longest baseline and most
complete time sampling.

Uncertainty in the intercalibration parameters for a given telescope
contributes to the final measurement uncertainty.  For a flux
measurement $f(t_{i})$ at epoch $t_i$, the calibrated measurement is
$f_c(t_i) = af(t_i) + b$, and standard error propagation shows that
the uncertainty introduced per point is $\sigma_{f_c}^2 =
\sigma_a^2f(t_i)^2 + \sigma_b^2 + 2f(t_i){\rm cov}(a,b)$.  Since $a$
and $b$ are usually anticorrelated, $\sigma_{f_c}$ is often small
compared to the uncertainties from image subtraction.  However, this
is not always the case for telescopes with very small numbers of
observations, so we calculated $\sigma_a$, $\sigma_b$, and ${\rm
  cov}(a,b)$ from the posterior distributions of these parameters for
each telescope, and added $\sigma_{f_c}$ in quadrature to the rescaled
{\tt ISIS} uncertainties for each epoch.  This treatment is very
conservative, since the intercalibration uncertainty is strongly
correlated between points from a single telescope.  A summary of the
mean intercalibration uncertainties is given in Table
\ref{tab:caldat}.

The choice of reference light curves defines the physical flux level
of the AGN from the corresponding {\tt ISIS} reference image (WC18 and
LT).  We convert the intercalibrated differential light curves to
physical flux units by performing aperture photometry on these
reference images.  For the AGN and all unsaturated stars in the field,
we measured the flux enclosed in a 5\farcs0 radius circular aperture,
and converted the summed fluxes to instrumental magnitudes.  This
means that the host-galaxy light within the aperture is included in
the measurement of the AGN flux, and this issue is discussed in \S2.4.
The background sky level was estimated from an annulus with
inner/outer radius of 14\arcsec/29\arcsec\ for the stars, and
118\arcsec/132\arcsec\ for the AGN (so as to avoid light from the host
galaxy).

\begin{figure*}
  \includegraphics[width=\textwidth]{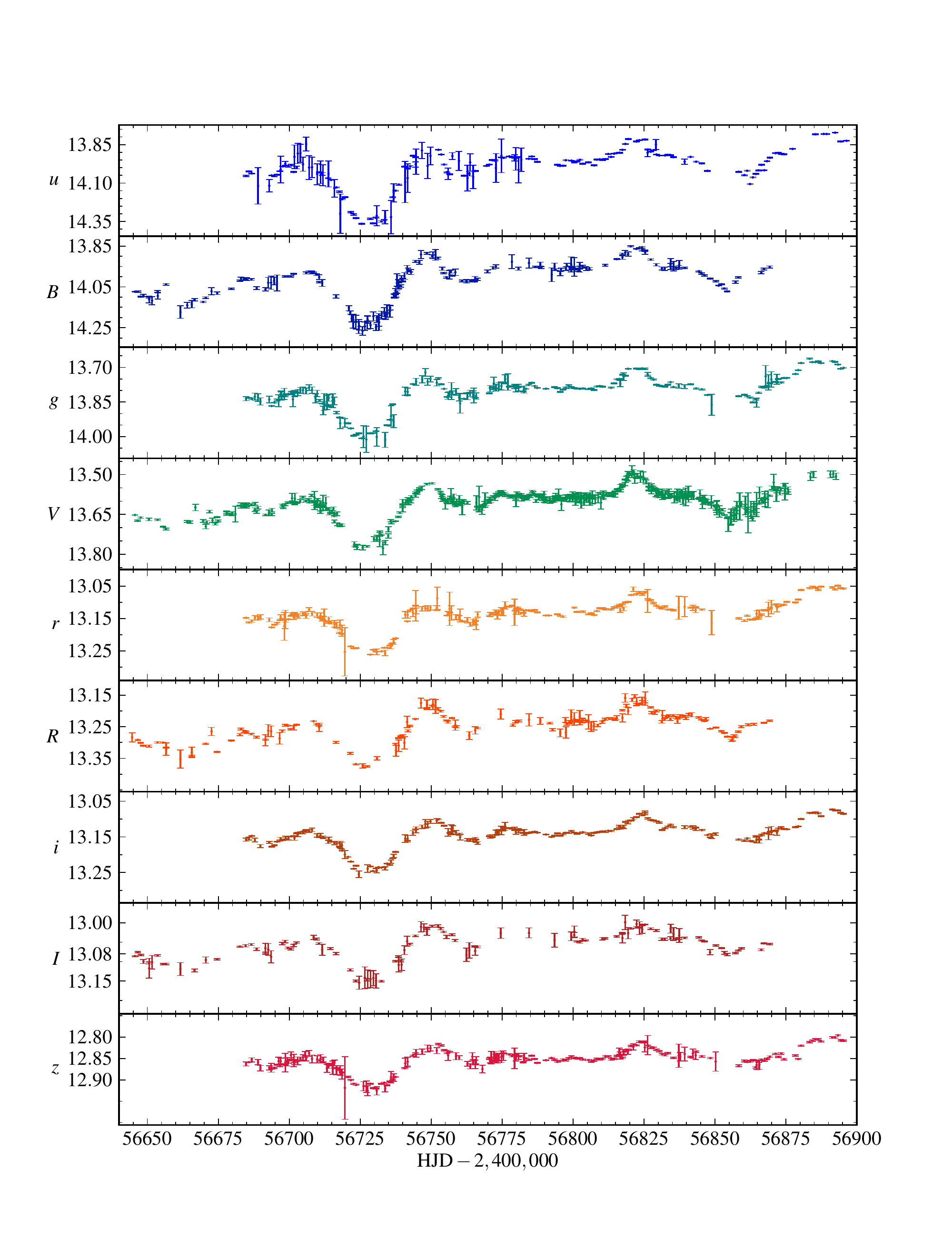}
  \caption{\emph{BVRI} and \emph{ugriz} ground-based light curves from
    the full monitoring campaign in AB magnitudes.  Only the
    measurement uncertainties in the differential fluxes are shown.
    These uncertainties include those due to intercalibration,
    summarized in Table \ref{tab:caldat}.  Systematic uncertainties
    for the absolute flux calibration are given in Table
    \ref{tab:lcprop}.  \label{fig:lc}}
\end{figure*}

We then matched all stars to the SDSS Data Release 7 catalog
\citep{Abazajian2007}, and computed the offset between instrumental
magnitudes and the SDSS AB magnitudes.  We did not find any
significant color terms in the flux calibration from the comparison
stars, although the spectral slope of the AGN would be a poor match to
such color terms regardless of their small values.  For the
Johnson/Cousins \emph{BVRI} bands, we determined the comparison-star
magnitudes using the filter-system transformations given by
\citet{Fukugita1996}, and converted these to AB magnitudes using
\citet{Fukugita1996} Table 8.  The filter transforms have an
uncertainty of $\sim\!  0.03$ mag, which we adopt as a floor for the
\emph{BVRI} flux-calibration uncertainty.  The final flux-calibrated
light curves are shown in Figure \ref{fig:lc} and given in Table
\ref{tab:lightcurves}.
\ifapj
  \begin{deluxetable}{lrrrrrrrrr}
\else
  \begin{deluxetable}{lrrrrrrrrr}
\fi
\tablecaption{Mean Intercalibration Uncertainty \label{tab:caldat}}
\tablewidth{0pt}
\tablehead{
\colhead{Telecope} & \colhead{{\it B}} & \colhead{{\it V}} & \colhead{{\it R}} & \colhead{{\it I}} & \colhead{{\it u}} & \colhead{{\it g}} & \colhead{{\it r}} & \colhead{{\it i}} & \colhead{{\it z}}\\
\colhead{(Matches Table \ref{tab:telescopes})} & \colhead{(\%)} & \colhead{(\%)} & \colhead{(\%)} & \colhead{(\%)} & \colhead{(\%)} & \colhead{(\%)} & \colhead{(\%)} & \colhead{(\%)} & \colhead{(\%)}
}
\startdata
WC18 & ref & ref & ref & ref & \dots & \dots & \dots & \dots & \dots \\
LT & \dots & \dots & \dots & \dots & ref & ref & ref & ref & ref \\
LCOGT1 & 0.9 & 0.2 & \dots & \dots & 1.1 & 1.2 & 0.5 & 0.1 & 0.3 \\
LCOGT2 & 2.3 & 0.4 & \dots & \dots & 1.8 & 0.5 & 1.9 & 0.4 & 1.9 \\
LCOGT3 & 1.5 & 0.6 & \dots & \dots & \dots & \dots & \dots & \dots & \dots \\
LCOGT4 & 2.7 & \dots & \dots & \dots & \dots & \dots & 0.3 & 0.1 & 0.2 \\
LCOGT5 & \dots & 0.6 & \dots & \dots & \dots & 1.3 & \dots & \dots & 0.7 \\
LCOGT6 & \dots & 0.5 & \dots & \dots & \dots & 1.6 & 1.4 & 0.4 & \dots \\
LCOGT7 & \dots & 1.9 & \dots & \dots & \dots & \dots & \dots & \dots & \dots \\
LCOGT8 & \dots & 1.3 & \dots & \dots & \dots & \dots & \dots & \dots & \dots \\
WMO & 0.9 & 0.6 & 0.3 & \dots & \dots & \dots & \dots & \dots & \dots \\
CrAO & 0.2 & \dots & 0.3 & 0.3 & \dots & \dots & \dots & \dots & \dots \\
RCT & 0.2 & 0.1 & \dots & \dots & \dots & \dots & \dots & \dots & \dots \\
MO15 & \dots & 0.4 & 0.7 & \dots & \dots & \dots & \dots & \dots & \dots \\
FWO & \dots & 0.3 & \dots & \dots & \dots & \dots & \dots & \dots & \dots \\
HLCO & \dots & 0.4 & \dots & \dots & \dots & \dots & \dots & \dots & \dots \\
KAIT & \dots & 0.1 & \dots & \dots & \dots & \dots & \dots & \dots & \dots \\
MLO & \dots & 0.7 & \dots & \dots & \dots & \dots & \dots & \dots & \dots \\
LOAO & \dots & 0.4 & \dots & \dots & \dots & \dots & \dots & \dots & \dots \\
\enddata
\tablecomments{Percentages are averaged for all epochs of the given
  telescope, measured relative to the flux at that epoch---see \S2.3
  for the definition of the intercalibration uncertainty.  ``ref'' is
  the reference telescope to which the others are aligned.}  \ifapj
  \end{deluxetable}
\else
  \end{deluxetable}
\fi

\ifapj
  \begin{deluxetable*}{lrcrrr}
\else
  \begin{deluxetable}{lrcrrr}
\fi

\tablecaption{Optical Continuum Light Curves \label{tab:lightcurves}}
\tablewidth{0pt} 
\tablehead{\colhead{Filter}  & \colhead{HJD} & \colhead{$F_{\lambda}$} & \colhead{Telescope ID} & \colhead{Differential Counts (DC)} & \colhead{error DC}\\
 &\colhead{ $-2,400,000$}&\colhead{($\rm 10^{-15}erg\,\,cm^{-2}\,s^{-1}\,$\AA$^{-1}$)} & \colhead{ (as in Table \ref{tab:telescopes})} & \colhead{ (reference counts)} & \colhead{ (reference counts)}
}
\startdata
{\it u} & 56684.78 & $ 21.61 \pm  0.08$ & LT & -34438.0 & 1292.5 \\
   & 56685.79 & $ 22.19 \pm  0.07$ & LT & -24994.0 & 1050.5 \\
   & 56686.77 & $ 21.93 \pm  0.04$ & LT & -29223.0 & 707.5 \\
  & \dots & \dots &\dots &\dots & \dots\\
{\it B}  & 56645.64 & $ 13.39 \pm  0.02$ & WC18 & -16159.0 & 460.55 \\
   & 56646.65 & $ 13.40 \pm  0.04$ & WC18 & -16055.0 & 764.35 \\
   & 56647.65 & $ 13.10 \pm  0.03$ & WC18 & -22095.0 & 595.81 \\
  & \dots & \dots &\dots &\dots & \dots\\
{\it g}  & 56684.78 & $ 13.96 \pm  0.12$ & LT & -8759.7 & 1384.1 \\
   & 56685.79 & $ 14.00 \pm  0.02$ & LT & -8305.6 & 192.33 \\
   & 56686.77 & $ 13.95 \pm  0.05$ & LT & -8892.6 & 579.47 \\
  & \dots & \dots &\dots &\dots & \dots\\
{\it V}  & 56645.62 & $ 12.91 \pm  0.02$ & WC18 & -4676.7 & 338.64 \\
   & 56646.61 & $ 12.66 \pm  0.03$ & WC18 & -8876.5 & 500.65 \\
   & 56647.63 & $ 12.79 \pm  0.03$ & WC18 & -6668.4 & 524.11 \\
  & \dots & \dots &\dots &\dots & \dots\\
{\it r}  & 56684.78 & $ 15.73 \pm  0.01$ & LT & -16832.0 & 491.06 \\
   & 56685.79 & $ 15.52 \pm  0.03$ & LT & -23984.0 & 915.05 \\
   & 56686.77 & $ 15.59 \pm  0.05$ & LT & -21698.0 & 1786.4 \\
  & \dots & \dots &\dots &\dots & \dots\\
{\it R}  & 56644.64 & $ 12.76 \pm  0.17$ & CrAO & -7835.7 & 4544.9 \\
   & 56646.63 & $ 12.67 \pm  0.02$ & WC18 & -10378.0 & 588.18 \\
   & 56647.64 & $ 12.56 \pm  0.02$ & WC18 & -13352.0 & 585.57 \\
  & \dots & \dots &\dots &\dots & \dots\\
{\it i}  & 56684.78 & $ 10.16 \pm  0.04$ & LT & -14639.0 & 2267.2 \\
   & 56685.78 & $ 10.19 \pm  0.04$ & LT & -13055.0 & 2074.4 \\
   & 56686.77 & $ 10.26 \pm  0.01$ & LT & -9307.6 & 458.99 \\
  & \dots & \dots &\dots &\dots & \dots\\
{\it I}  & 56645.63 & $ 8.67 \pm  0.01$ & WC18 & -6575.6 & 499.37 \\
   & 56646.63 & $ 8.75 \pm  0.02$ & WC18 & -3043.8 & 732.31 \\
   & 56647.64 & $ 8.69 \pm  0.02$ & WC18 & -5582.9 & 702.53 \\
  & \dots & \dots &\dots &\dots & \dots\\
{\it z}  & 56684.78 & $ 9.30 \pm  0.04$ & LT & -4697.2 & 984.0 \\
   & 56685.78 & $ 9.34 \pm  0.01$ & LT & -3818.5 & 197.04 \\
   & 56686.77 & $ 9.38 \pm  0.04$ & LT & -2659.7 & 1013.4 \\
  & \dots & \dots &\dots &\dots & \dots\\
\enddata
\tablecomments{This table is available in its entirety in the online version of this article.}
\ifapj
  \end{deluxetable*}
\else
  \end{deluxetable}
\fi

\ifapj
  \begin{deluxetable}{ccc}
\else
  \begin{deluxetable}{ccc}
\fi
\tablecaption{{\it HST}  Continuum Light Curves \label{tab:HST_lightcurves}}
\tablewidth{0pt} 
\tablehead{ \colhead{Wavelength} & \colhead{HJD} & \colhead{$F_{\lambda}$}\\
\colhead{(\AA)} &\colhead{ $-2,400,000$}&\colhead{($\rm 10^{-15}erg\,\,cm^{-2}\,s^{-1}\,$\AA$^{-1}$)
}
}
\startdata
1157.5 & 56690.61 & $32.40 \pm 0.89$ \\
\dots & 56691.54 & $34.80 \pm 0.92$ \\
\dots & 56692.39 & $37.50 \pm 0.95$ \\
\dots & \dots     & \dots\\
1367. & 56690.61 & $34.27 \pm 0.64$ \\
\dots & 56691.54 & $35.45 \pm 0.65$ \\
\dots & 56692.39 & $37.71 \pm 0.67$ \\
\dots & \dots     & \dots\\
1478.5 & 56690.65 & $29.70 \pm 0.48$ \\
\dots & 56691.58 & $31.60 \pm 0.51$ \\
\dots & 56692.41 & $33.00 \pm 0.52$ \\
\dots & \dots     & \dots\\
1746. & 56690.65 & $26.70 \pm 0.63$ \\
\dots & 56691.58 & $27.90 \pm 0.64$ \\
\dots & 56692.41 & $30.40 \pm 0.68$ \\
\enddata
\tablecomments{This table is available in its entirety in the online version of this article.}
\ifapj
  \end{deluxetable}
\else
  \end{deluxetable}
\fi

\subsection{Light-Curve Properties}
\begin{figure*}
\includegraphics[width=1.0\textwidth]{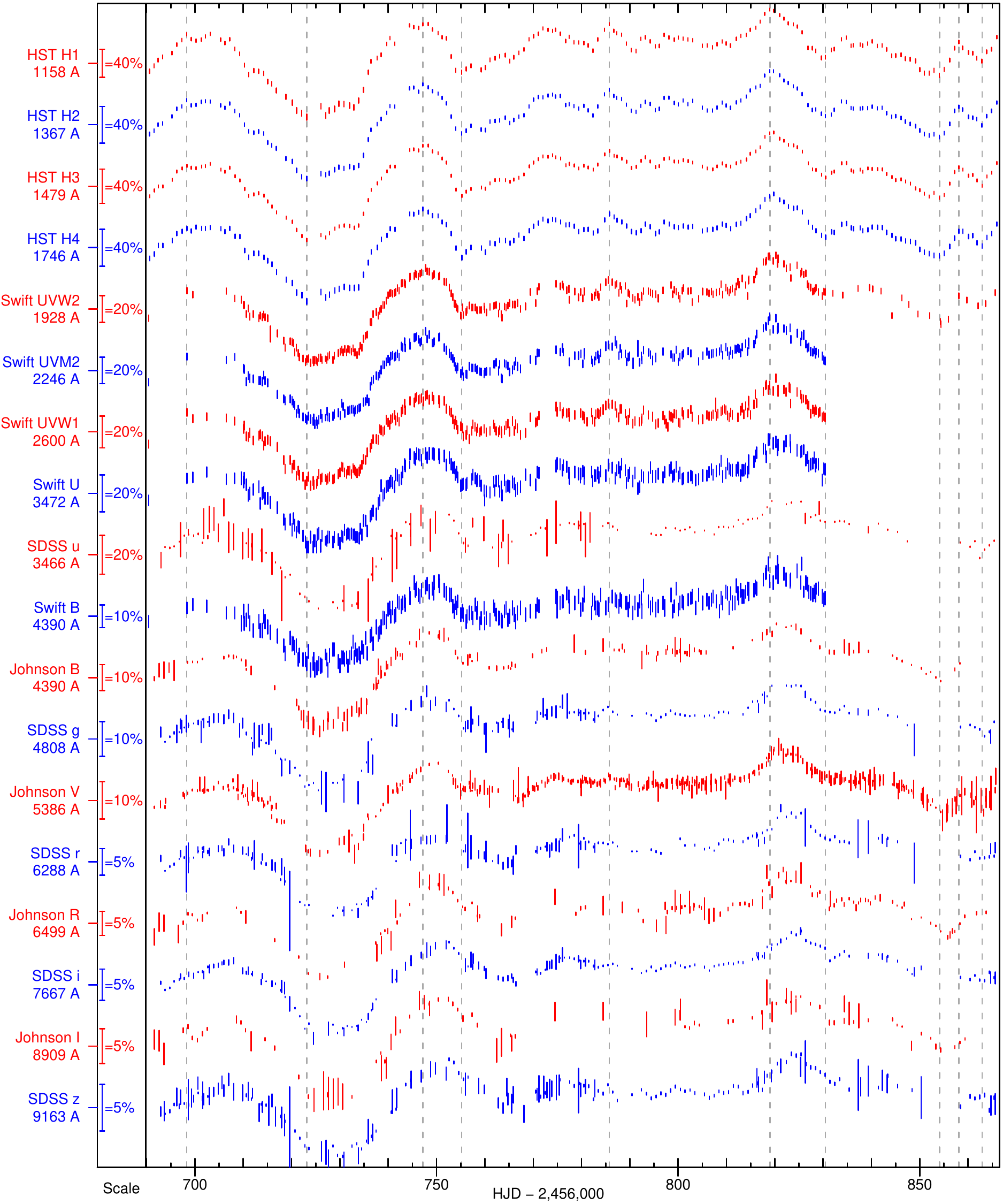}
\caption{AGN STORM UV and optical continuum light curves used in this
  analysis, restricted to the observing window of the {\it HST}
  campaign.  Light curves have been converted to AB magnitudes, but are
  rescaled and shifted for clarity---the scales along the vertical
  axis show the fractional variations.  The vertical dashed lines mark
  local extrema in the {\it HST} 1158\,\AA\
  light curve.  \label{fig:all_lc}}
\end{figure*}

Table \ref{tab:lcprop} gives a summary of the sampling properties of
the AGN STORM continuum light curves, and shows that the light curves
have approximately daily cadence over the entire campaign.  Paper I
and Paper II only presented the \emph{HST} \drivelc\ continuum light
curve; here, we include three additional UV continuum light curves
measured from the \emph{HST} COS spectra, extracted from 5--6\,\AA\
windows centered at 1158\,\AA, 1479\,\AA, and 1746\,\AA, and given in Table \ref{tab:HST_lightcurves}.  These
continuum windows were chosen to be as uncontaminated as possible by
absorption lines and broad emission-line wings.  We also drop the {\it
  Swift V}-band light curve from this analysis, because its mean
fractional uncertainty is much larger than that of the ground-based
Johnson {\it V}-band light curve (3.2\% compared to 0.8\%).  The
reported wavelengths of the optical light curves are pivot wavelengths
calculated from the filter response curves of the optical bands, and
they are independent of the source spectrum (atmospheric cutoffs at
3000\,\AA\ and 1\,$\mu$m were imposed for these calculations).  Figure
\ref{fig:all_lc} shows a comparison of all the continuum light curves
used in this study.

Table \ref{tab:lcprop} also gives the variability properties of the light 
curves. Column 8 gives the mean flux and root-mean square (rms) scatter of 
the light curves, corrected for Galactic extinction assuming a
\citet*{Cardelli1989} extinction law with $R_V=3.1$ and $E(B-V) =
0.0171$\ mag (\citealt{Schlegel1998,Schlafly2011}; Paper I).  Columns
10, 11, and 12 give different estimates of their fractional
variability.  The fractional variability $F_{\rm var}$ of a light
curve is defined by
\begin{align}
  F_{\rm var} = \frac{1}{{\langle f(t) \rangle}} \sqrt{\frac{1}{N}\sum_i^N   \left\{\left[f(t_i) - \langle f(t) \rangle \right]^2 - \sigma_{i}^2  \right\}  }
\end{align}
and the uncertainty in $F_{\rm var}$ is
 \begin{align}
\sigma_{F_{\rm var}}^2 = \left(
  \sqrt{\frac{1}{2N}} 
  \frac{\langle \sigma^2\rangle}{\langle f(t)\rangle^2 F_{\rm var}}
\right)^2 +
 \left(
\sqrt{\frac{\langle \sigma^2\rangle}{N}} 
\frac{1}{\langle f(t) \rangle}
\right)^2,
\end{align}
where $f(t_i)$ is the value of the light curve at epoch ${i}$,
$\sigma_i$ is the associated uncertainty, $\langle f(t)\rangle$ is the
(unweighted) mean value of the light curve, and $\langle \sigma^2
\rangle$ is the mean square of the measurement uncertainties
\citep{Rodriguez1997,Vaughan2003}.  We also estimated the fractional
variability using the {\tt JAVELIN} amplitudes, $\sigma_{\rm
  DRW}/\langle f \rangle$, since this is an equivalent measure of
$F_{\rm var}$ under the DRW model.  The values of $\sigma_{\rm
  DRW}/\langle f \rangle$ and $F_{\rm var}$ are often in good
agreement, but with notable exceptions, as given in Table
\ref{tab:lcprop}.

\begin{figure*}
  \includegraphics[width=\textwidth]{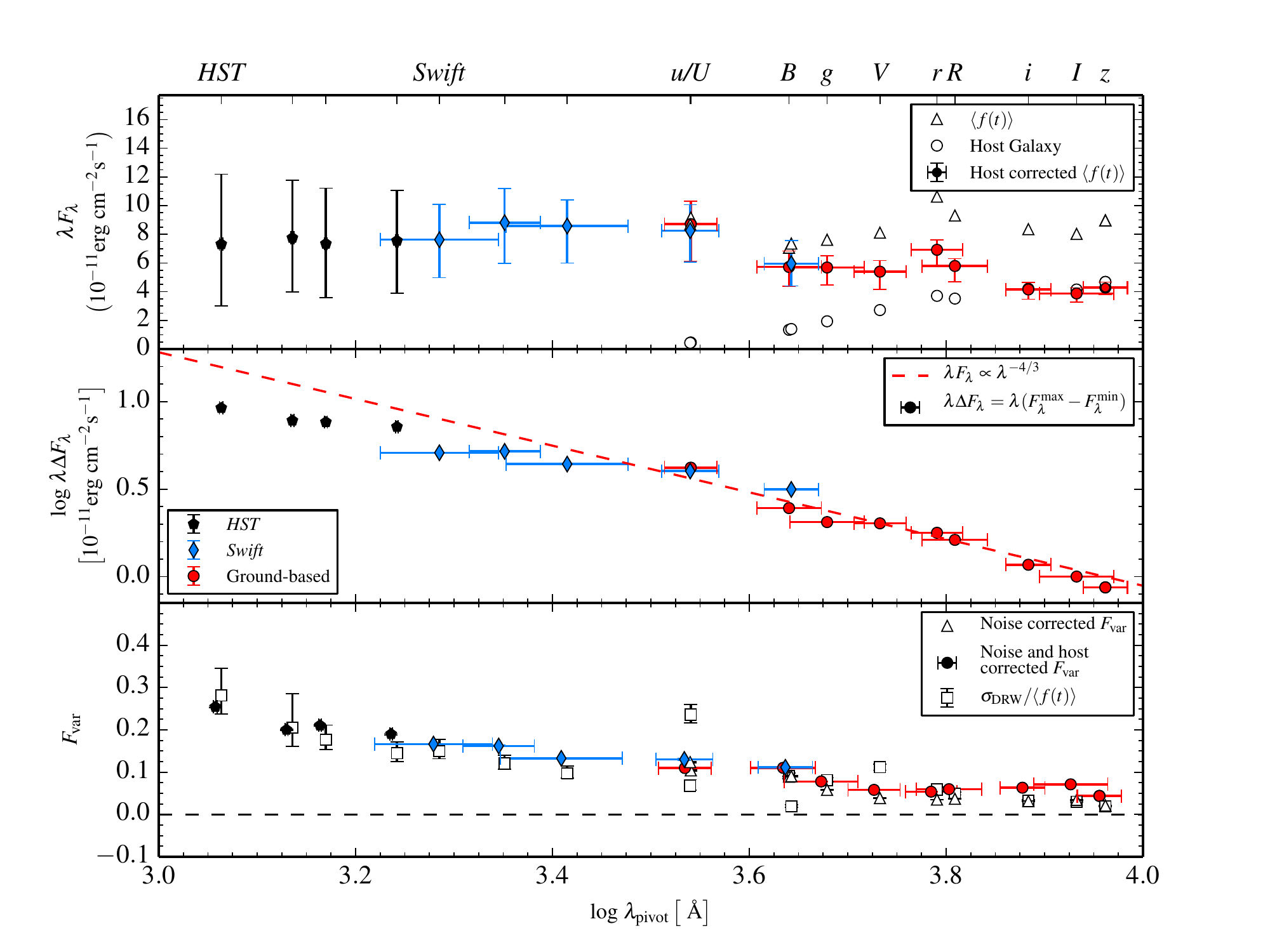}
  \caption{{\bf Top Panel:} Mean SED of NGC 5548 from far-UV to
    optical wavelengths, corrected for Galactic extinction.  The
    vertical error bars represent the AGN in the maximum and minimum
    states of the campaign.  The horizontal error bars represent the
    rms width of the filter transmission curves.  See \S4 for a
    discussion of the host-galaxy estimate.  {\bf Middle Panel:}
    Variable SED component, calculated from the difference in flux
    between the minimum and maximum states, which more cleanly
    identifies the accretion-disk spectrum.  The dashed red line is
    the predicted spectrum for a standard thin disk---discrepancies at
    short wavelengths may be due to extinction internal to the AGN or
    the inner edge of the disk. {\bf Bottom Panel:} Fractional
    variability $F_{\rm var}$ as a function of wavelength.
    $\sigma_{\rm DRW}$ is the DRW amplitude from the {\tt JAVELIN}
    fits.  For clarity, a small shift in wavelength to the $F_{\rm
      var}$ points has been applied.\label{fig:F_var} }
\end{figure*}

Figure \ref{fig:F_var} shows the mean flux and variability
properties of these light curves.  The top panel displays the mean SED
(corrected for Galactic extinction).  The vertical error bars show the
minimum and maximum states of the AGN, which occur at ${\rm HJD} -
2,400,000 = 56,723.1$ and $56,818.9$, respectively.  These dates are based
on the {\it HST} 1367\,\AA\ light curve, and the other bands are
adjusted for interband time delays that are measured in \S3.  The
middle panel illustrates the logarithm of the difference in flux between the
minimum and maximum states of the AGN, which cleanly isolates the
variable component of the spectrum and better traces the shape of the
accretion-disk SED.  For comparison, a standard thin accretion disk
SED with $\lambda F_{\lambda}\propto \lambda^{-4/3}$ is shown,
arbitrarily normalized to match the Johnson {\it V}-band differential
flux.  Although the data are in excellent agreement with the
prediction at longer wavelengths, the UV data lie significantly below
the model SED.  This discrepancy may be caused by extinction internal to
the AGN, or the inner edge of the disk, which will display an
exponential Wien cutoff rather than $\lambda F_{\lambda}\propto
\lambda^{-4/3}$.  A more complete discussion and modeling of the
variable spectrum will be presented by Starkey et al. (in prep.).
Finally, the bottom panel shows $F_{\rm var}$ as a function of
wavelength. The far-UV light curves have values of $F_{\rm var} \geq
0.20$, which sharply decrease with wavelength to about 0.06 in the
{\it V} band.  At longer wavelengths, the trend flattens, reaching
0.02 in the {\it z} band.

At least part of this effect is caused by the constant flux contributed
by the host galaxy, which becomes increasingly important at longer
wavelengths.  Based on spectral decomposition models and synthetic
photometry (described in \S4.1 and \S4.2), the host galaxy contributes
about 20\% of the observed flux in the \emph{B} band, and about 54\%
in the \emph{I} and \emph{z} bands.  We corrected $\langle f(t)
\rangle $ for this constant component, and Figure \ref{fig:F_var} and
Table \ref{tab:lcprop} also show the host-galaxy flux and revised
values of $F_{\rm var}$. The effect on the trend in Figure
\ref{fig:F_var} is fairly subtle, and does not change the flattening
at optical wavelengths.

The larger variability amplitudes at short wavelengths suggest that the
SED of NGC 5548 becomes bluer in higher flux states.  The same effect
was seen by \citet{Cackett2015} in NGC 5548 with near-UV grism
monitoring data from {\it Swift}.  However, the trend is driven by the
light curves at wavelengths $<\!  5000$\,\AA, and is most significant
at wavelengths $\lesssim 3500$\,\AA, which may be why optical studies
of AGN variability do not always find any ``bluer when brighter''
trend \citep[e.g.,][]{Sakata2010}.
\ifapj
  \begin{deluxetable*}{lrrcrrrrcrrr}
\else
  \begin{deluxetable}{lrrcrrrrcrrr}
  \rotate
    \fi \tablecaption{Light Curves Properties \label{tab:lcprop}}
    \tablewidth{0pt} \tablehead{
      \colhead{Source} & \colhead{Filter} & \colhead{$\lambda_{\rm pivot}$}& \colhead{Flux Calibration} & \colhead{$N_{\rm obs}$} & \colhead{$\Delta t_{\rm ave}$} & \colhead{$\Delta t_{\rm med}$} & \colhead{$\langle f(t) \rangle$\tablenotemark{a}} & \colhead{Host\tablenotemark{a}} & \colhead{$F_{\rm var}$} & \colhead{$F_{\rm var2}$\tablenotemark{b}} & \colhead{$\sigma_{\rm DRW}/\langle f(t) \rangle$}\\
      & & \colhead{(\AA)} & \colhead{Uncertainty (mag)} &
      &\colhead{(days)}& \colhead{(days)}\\
\colhead{(1)}&\colhead{(2)}&\colhead{(3)}&\colhead{(4)}&\colhead{(5)}&\colhead{(6)}&\colhead{(7)}&\colhead{(8)}&\colhead{(9)}&\colhead{(10)}&
\colhead{(11)}&\colhead{(12)}
 }

\startdata
\emph{HST} & $\lambda 1158$ & 1158 & 0.050 & 171 & 1.03 & 1.00 & $52.41\pm 13.38$ & \dots & $ 0.254\pm 0.002$ & $ 0.254\pm 0.002$ & $  0.281 \pm  0.054$ \\
\emph{HST} & $\lambda 1367$ & 1367 & 0.050 & 171 & 1.03 & 1.00 & $49.17\pm 9.89$  & \dots & $ 0.200\pm 0.001$ & $ 0.200\pm 0.001$ & $  0.205 \pm  0.062$ \\
\emph{HST} & $\lambda 1479$ & 1479 & 0.050 & 171 & 1.03 & 1.00 & $43.54\pm 9.20$  & \dots & $ 0.211\pm 0.001$ & $ 0.211\pm 0.001$ & $  0.176 \pm  0.029$ \\
\emph{HST} & $\lambda 1746$ & 1746 & 0.058 & 171 & 1.03 & 1.00 & $38.26\pm 7.32$  & \dots & $ 0.190\pm 0.002$ & $ 0.190\pm 0.002$ & $  0.145 \pm  0.024$ \\
\emph{Swift} & {\it UVW2} & 1928 & 0.030 & 284 & 0.62 & 0.39 & $34.71\pm 5.83$    & \dots & $ 0.166\pm 0.001$ & $ 0.166\pm 0.001$ & $  0.150 \pm  0.023$ \\
\emph{Swift} & {\it UVM2} & 2246 & 0.030 & 256 & 0.55 & 0.35 & $33.83\pm 5.55$    & \dots & $ 0.162\pm 0.002$ & $ 0.162\pm 0.002$ & $  0.121 \pm  0.017$ \\
\emph{Swift} & {\it UVW1} & 2600 & 0.030 & 270 & 0.52 & 0.38 & $29.70\pm 4.01$    & \dots & $ 0.133\pm 0.002$ & $ 0.133\pm 0.002$ & $  0.097 \pm  0.014$ \\
\emph{Swift} & {\it U} & 3467 & 0.020 & 145 & 1.20 & 0.99 & $24.43\pm 2.59$ & $1.22\pm 0.02$ & $ 0.104\pm 0.002$ & $ 0.110\pm 0.002$ & $  0.236 \pm  0.021$ \\
Ground & {\it u} & 3472 & 0.035 & 270 & 0.52 & 0.38 & $23.18\pm 2.94$ & $1.16\pm 0.02$ & $ 0.124\pm 0.002$ & $ 0.130\pm 0.002$ & $  0.068 \pm  0.012$ \\
Ground & {\it B} & 4369 & 0.030 & 151 & 1.11 & 0.98 & $15.15\pm 1.36$ & $2.88\pm 0.05$ & $ 0.089\pm 0.001$ & $ 0.110\pm 0.001$ & $  0.090 \pm  0.007$ \\
\emph{Swift} & {\it B Swift} & 4392 & 0.016 & 271 & 0.52 & 0.37 & $15.69\pm 1.48$ & $2.98\pm 0.05$ & $ 0.090\pm 0.002$ & $ 0.112\pm 0.002$ & $  0.019 \pm  0.003$ \\
Ground & {\it g} & 4776 & 0.034 & 172 & 1.01 & 0.97 & $15.06\pm 0.89$ & $3.83\pm 0.08$ & $ 0.058\pm 0.001$ & $ 0.078\pm 0.001$ & $  0.081 \pm  0.005$ \\
Ground & {\it V} & 5404 & 0.030 & 429 & 0.41 & 0.31 & $14.29\pm 0.56$ & $4.79\pm 0.10$ & $ 0.039\pm 0.001$ & $ 0.058\pm 0.001$ & $  0.112 \pm  0.007$ \\
Ground & {\it r} & 6176 & 0.032 & 172 & 1.01 & 0.93 & $16.49\pm 0.59$ & $5.76\pm 0.12$ & $ 0.035\pm 0.001$ & $ 0.054\pm 0.001$ & $  0.059 \pm  0.005$ \\
Ground & {\it R} & 6440 & 0.030 & 136 & 1.28 & 0.96 & $13.88\pm 0.52$ & $5.25\pm 0.10$ & $ 0.037\pm 0.001$ & $ 0.060\pm 0.001$ & $  0.049 \pm  0.003$ \\
Ground & {\it i} & 7648 & 0.021 & 178 & 0.98 & 0.96 & $10.59\pm 0.33$ & $5.33\pm 0.10$ & $ 0.031\pm 0.001$ & $ 0.063\pm 0.001$ & $  0.032 \pm  0.002$ \\
Ground & {\it I} & 8561 & 0.030 & 98 & 1.73 & 1.02 & $9.15\pm 0.32$ & $4.73\pm 0.08$ & $ 0.034\pm 0.001$ & $ 0.071\pm 0.001$ & $  0.030 \pm  0.002$ \\
Ground & {\it z} & 9157 & 0.011 & 186 & 0.93 & 0.91 & $9.57\pm 0.21$ & $5.00\pm 0.08$ & $ 0.021\pm 0.001$ & $ 0.044\pm 0.001$ & $  0.019 \pm  0.002$ \\
\enddata
    \tablecomments{$N_{\rm obs}$ gives the number of epochs in the lightcurve,
      $\Delta t_{\rm ave}$ gives the average cadence, $\Delta t_{\rm
        med}$ gives the median cadence, $\langle f(t) \rangle$ gives
      the mean flux (the uncertainty gives the rms scatter of the
      lightcurve), ``Host'' gives the host-galaxy flux, $F_{\rm var}$
      is defined in \S2.4, and $\sigma_{\rm DRW}$ is the DRW
      amplitude.  The flux calibration uncertainty is the systematic
      uncertainty for conversion to physical units (i.e., zeropoint
      errors).  For {\it HST}, these values are taken from Paper I,
      while for {\it Swift} the values are from Table 6 of
      \cite{Poole2008}. The uncertainties for the ground-based
      lightcurves represent our calibration to the SDSS AB mag
      photometric system.  A correction for Galactic extinction has
      been applied to these data (see \S2.4 for details).}

\tablenotetext{a}{$\rm 10^{-15} erg\,\,cm^{-2}\,s^{-1}$\AA$^{-1}$}
\tablenotetext{b}{Corrected for host-galaxy flux}

 \ifapj
  \end{deluxetable*}
\else
  \end{deluxetable}
\fi

\section{Time-Series Analysis}
We measure the lags between light curves using two methods.  First, we
use the interpolated cross-correlation function (ICCF), as employed by
\citet{Peterson2004}, and estimate the uncertainty of the lag using a
Monte Carlo method.  Second, we use {\tt JAVELIN}, which measures lags
by modeling reverberating light curves as shifted, scaled, and
smoothed versions of the driving light curve.

In the first method, the ICCF is calculated by shifting one light
curve on a grid of lags $\tau$ spaced by 0.1 day, and calculating the
correlation coefficient $r_{cc}(\tau)$ by linearly interpolating the
second light curve.  The lags are estimated from the centroid of the
ICCF, defined as the mean ICCF-weighted lag for which $r_{cc}(\tau) >
0.8\,r_{\rm max}$.  Uncertainties are estimated using the flux
randomization/random subset selection (FR/RSS) method, wherein a
distribution of ICCF centroids is built from cross correlating $10^3$
realizations of both light curves.  Each realization consists of
randomly selected epochs (chosen with replacement), and the
corresponding flux measurements are adjusted by random Gaussian
deviates scaled to the measurement uncertainties.  The lags reported
here correspond to the medians of the ICCF centroid distributions,
while the lower and upper uncertainties define their central 68\%
confidence intervals.

\begin{figure*}
  \includegraphics[width=\textwidth]{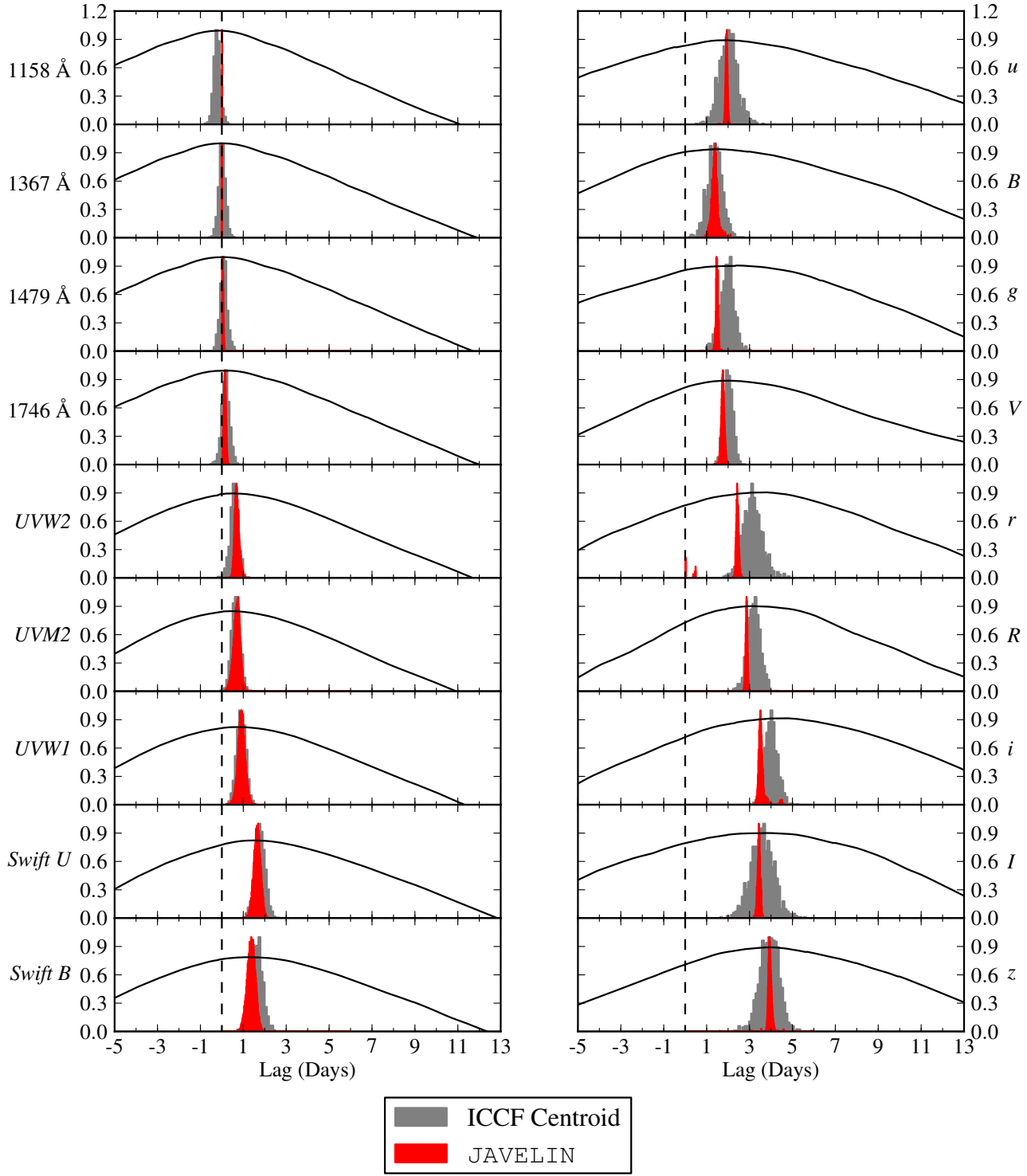}
  \caption{ICCF for all light curves, with the ordinate showing the
    correlation coefficient $r_{cc}$.  Lags for data from Paper I and
    Paper II (following our reanalysis) are shown in the left column,
    ground-based optical lags are presented in the right column.  The
    grey histograms are the ICCF centroid distributions from the
    FR/RSS method, the red histograms are from {\tt JAVELIN}.  Both
    histograms are in units of $P(\tau)/{\rm max}[P(\tau)]$.\label{fig:example_dist} }
\end{figure*}

We detrended the light curves, as is common practice in RM studies
(\citealt{Peterson2004}; Paper II), in order to remove long-term
secular trends that are poorly sampled in the frequency domain and may
bias the observed lag \citep{Welsh1999}.  The detrending procedure
consists of subtracting a second-order polynomial linear least-squares
fit (with equal weight given to all data points) from the observed
light curves.  Following Paper I and Paper II, we restricted the
analysis to the time period coincident with the \emph{HST} campaign,
and measured the time delays relative to the \emph{HST} \drivelc\
light curve.  When calculating the ICCF, we only interpolate the
\drivelc\ light curve.  Table \ref{tab:lags} summarizes the resulting
mean lags, corrected for cosmological time dilation (the redshift of
NGC 5548 is $z = 0.017175$; Paper I).  Lags for the hard and soft
X-ray bands of the {\it Swift} XRT are also included, as determined in
Paper II.  The ICCF for all bands is shown in Figure
\ref{fig:example_dist} with the solid black lines, while the centroid
distributions are shown as the gray histograms.  We found that the
{\it HST} 1158\,\AA\ and 1479\,\AA\ lags were only slightly larger
than the spacing of our interpolation grid (0.1 day), so we repeated
the procedure on these light curves using a grid of 0.01 day.  This
did not have a noticeable effect on the ICCF centroids, but it did
change the ICCF peaks by $\sim\!  0.05$ day.  The lags reported in
Table \ref{tab:lags} make use of the finer grid for these light
curves.

Our treatment of the \emph{Swift} light curves (\emph{UVW2, UVM2,
  UVW1, U, {\rm and} B}) results in lags systematically larger than
those found in Paper II, although the tension is only moderate
(typically $\lesssim\! 1.5\sigma$).  These differences are primarily
caused by the different detrending procedures of the two studies.  Paper
II detrended the \emph{Swift} light curves using a 30-day running
mean, while we use a low-order polynomial. A running mean corresponds
to a lower-pass filter than a polynomial trend, so this detrending
procedure removes more low-frequency power from the light curve and
is therefore expected to result in smaller lags.  However, several of
our light curves have very irregular sampling, which makes the
calculation of the running mean poorly defined, so we instead use the
low-order polynomial.  The ground-based SDSS {\it u} and {\it Swift U}
lags and the Johnson {\it B} and {\it Swift B} lags are consistent at
the $\sim\!  0.6\sigma$ level using the polynomial detrending, so it
is likely that the detrending procedure accounts for most of the
difference between the near-UV lags.  Two other smaller effects may be
important for the lag determinations.  First, the \emph{Swift} UVOT
optical filters are much narrower than the standard Bessell filters
used for the ground-based light curves, so the observed variations are
not perfectly identical (the \emph{Swift} light curves also have
slightly shorter baselines).  Second, the \emph{Swift} optical light
curves have much larger fractional uncertainties, which may shift the
ICCF centroid distribution of the otherwise similar light curves.

We also estimate the lags using {\tt JAVELIN}, which calculates a
maximum-likelihood lag, scale factor, and kernel width (assuming a
top-hat transfer function) from the DRW covariance matrices.  {\tt
  JAVELIN} internally employs a linear detrending procedure, so we do
not apply the second-order detrending as for the ICCF analysis.  We
also imposed a minimum kernel width of 0.75 day, in order to suppress
solutions where {\tt JAVELIN} finds a $\delta$-function transfer
function and aligns the reverberating light curve with the gaps
between samples of the driving light curve (this is an aliasing
problem associated with light curves that have similar cadences).

We adopt the medians of the posterior lag distributions and their
central 68\% confidence intervals as estimates of the lag and its
uncertainty, which are given in Table \ref{tab:lags}.  The posterior
distributions are shown by the red histograms in Figure
\ref{fig:example_dist}.  The median lags are always consistent with
the ICCF analysis, with the largest discrepancy being 1.7$\sigma$ in
the {\it r} band.  The {\tt Javelin} uncertainties generically appear
to be uncomfortably small.  This is because {\tt JAVELIN} assumes
correctly characterized random Gaussian measurement errors, that the
line light curve is a simple lagged and smoothed version of the
continuum light curve, and that the smoothing kernel is well
characterized by the functional form of the model (a top-hat
function).  Given that all these requirements are seldom fully met
(particularly the Gaussianity of the measurement errors), {\tt
  Javelin} uncertainties need to be interpreted conservatively.  A
rough rule of thumb from modeling gravitational lens time delays is
that repeated measurements for the same system will typically be
within 2$-$3$\sigma$ of each other.

The very small {\tt JAVELIN} uncertainties may also indicate that the
simple lagged and smoothed model of the reverberating light-curve
model is an inadequate description of the data.  Paper I found a
similar result, where the shape of the line light curves was not
always a good match to the observed continuum light curve.  Therefore,
smoothing the continuum light curve by a simple transfer function
cannot always reproduce the line light curve, suggesting that other
processes are important for the observed line emission (perhaps, for
example, anisotropic emission/reprocessing).  A more detailed
investigation of this result will be pursued in upcoming papers of
this series.

\begin{figure*}
  \includegraphics[width=\textwidth]{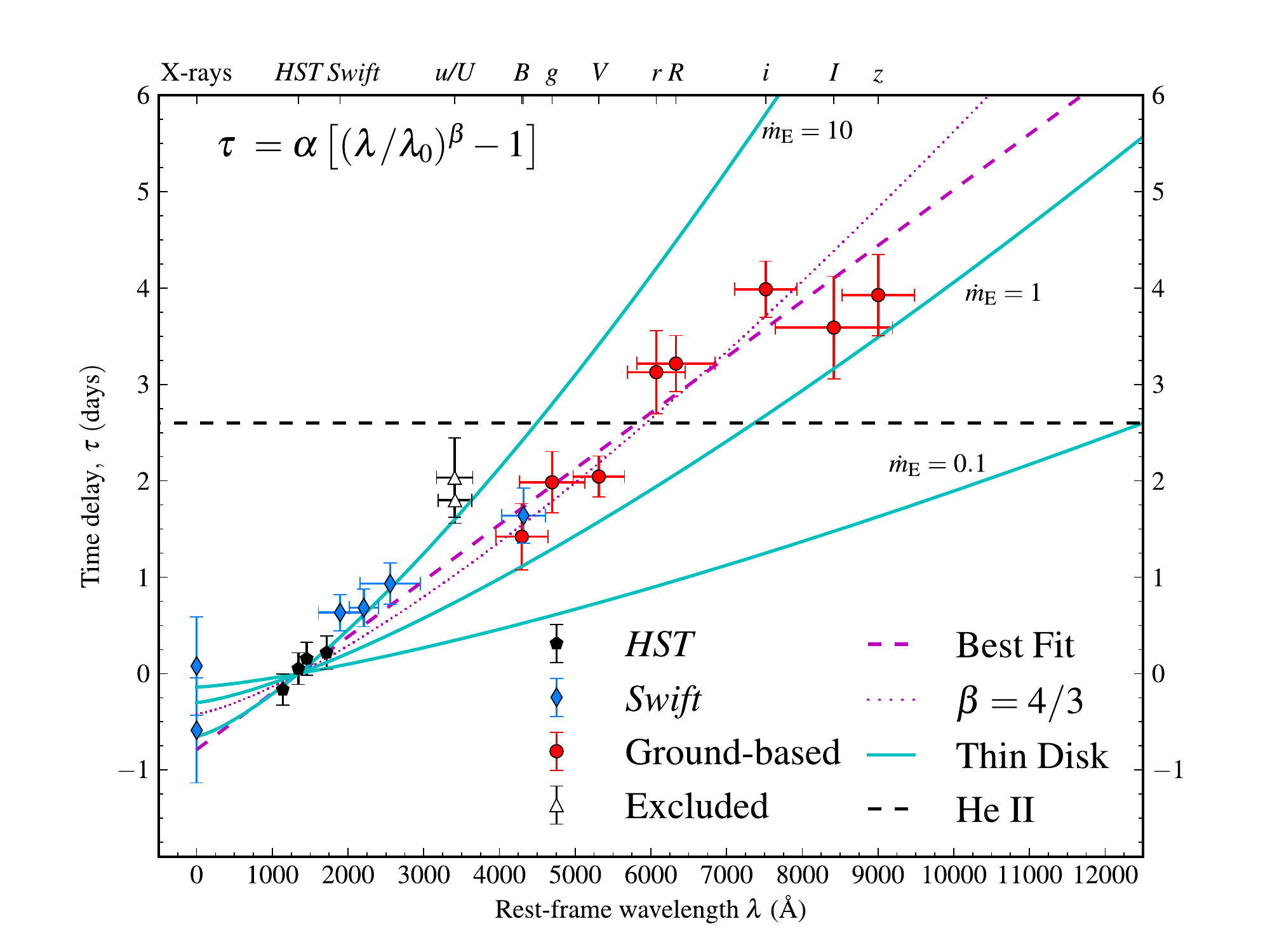}
  \caption{Time delay (ICCF centroid) as a function of pivot
    wavelength of the filters.  The horizontal error bars represent
    the rms width of the filters.  The best-fit model is shown by the
    dashed magenta line, while the fit fixing $\beta = 4/3$ is shown
    by the dotted magenta line.  Predictions for a thin-disk model
    with $\dot m_{\rm E} = L / L_{\rm Edd}$ are shown by the solid
    cyan lines, although the assumptions of the model are unlikely to
    hold at large $\dot m_{\rm E}$ (see \S5.3).  The mean lag of the
    He\,{\sc ii}$\,\lambda1640$ and $\lambda 4686$ lines is shown by
    the horizontal dashed black line (Paper I,
    \citePei).  \label{fig:lag_wv} }
\end{figure*}

Using either lag estimation technique, we find that longer wavelength
continuum variations follow those at shorter wavelengths.  Figure
\ref{fig:lag_wv} shows the lags as a function of the pivot wavelength
of each filter.  While the far-UV and near-UV light curves have time
delays $\tau < 1$ day, the \emph{V} band lags the \drivelc\ continuum
by \Vlag\ days, and the \emph{z} band lags it by $3.93 \pm 0.42$ days.
For comparison, the He\,{\sc ii} UV and optical lines (1640\,\AA\ and
4686\,\AA, respectively) have a mean lag of $\sim\!  2.5$ days relative to the
\drivelc\ light curve (Paper I, \citePei).  The optical light curves
have a time delay comparable to, and frequently larger than, that of
high-ionization-state lines in the BLR.

The trend of larger lags at longer wavelengths is nearly monotonic.
The most notable exceptions are (1) in the longest-wavelength filters,
where the trend appears to level out near the \emph{i} band, and (2)
in the \emph{u {\rm and} U} bands.  The \emph{u {\rm and} U} bands
have mean lags of $2.03\pm 0.41$ days and $1.80\pm 0.24$ days,
respectively, comparable to or larger than the lags of the \emph{g
  {\rm and} V}-band light curves. This may be due to emission
originating in the BLR picked up in the \emph{u {\rm and} U}-band
filters, which would contaminate measurements of the AGN continuum
emission and artificially increase the observed lag.  A similar
explanation may exist for the downturn at the {\it I} and {\it z}
bands, since Paschen continuum emission from the BLR begins at
8204\,\AA\ (see \citealt{Korista2001}).  We return to the question of
BLR contamination in \S4.

 Optical continuum lags in NGC 5548 have previously been measured by
\citet{Sergeev2005}, and the same light curves were reexamined by
\citet{Chelouche2013a} and \citet{Chelouche2013b}.
\citet{Sergeev2005} found substantially longer time delays between the
{\it B} and {\it R}/Cousins {\it I} bands than the lags presented here
(about 8 days).  However, the \citet{Sergeev2005} light curves have
$\sim\! 3$\ day cadence and suffer from large seasonal/scheduling gaps
of 20 days or more.  The difference in the optical lags is therefore
likely caused by systematic issues with the \citet{Sergeev2005} light
curves, such as unfortunate gaps that affect the cross-correlation
functions.  On the other hand, \citet{Chelouche2013a} and
\citet{Chelouche2013b} claim that the large optical lags are due to
BLR contamination and that the true continuum lags are consistent with
zero.  We discuss this possibility further in \S4.3, but we find this
interpretation to be unlikely.  These studies did not discuss the
impact of gaps in the data on the multivariate cross-correlation
function used to disentangle line and continuum lags, and we are
further skeptical that this method can meaningfully measure lags below
the cadence of the light curves (3 days, in this case).

To avoid the systematics associated with small lags, interband
continuum lags should be measured with data taken near or well below
the timescale of any suspected lags.  The UV wavelength coverage of
the STORM project therefore lends a tremendous boost to our ability to
detect the continuum lags, since the UV-optical lags are 3 to 6 times
larger than the interband optical lags.  This has implications for
ground-based studies attempting to resolve interband continuum lags.
Since the optical lags are of order 1 day (or less), the diurnal cycle
may make it impossible to measure reliable interband optical lags
from the ground without favorable conditions.

In order to quantify the trend of lag with wavelength, we fit a model
to the data presented in Figure \ref{fig:lag_wv} using the functional
form
\begin{align}
\label{equ:lag_wv}
\tau = \alpha\left[\left(\frac{\lambda}{\lambda_0} \right)^{\beta}
  -1\right],
\end{align}
where $\tau$ is the observed lag, $\lambda_0$ is a reference
wavelength, and $\alpha$ and $\beta$ are free parameters.  As in Paper
II, we set $\lambda_0 = 1367$\,\AA\ and report all covariances between
parameters.  The results of the fits are summarized in Table
\ref{tab:fitparams}.  We find that $\alpha =0.97\pm 0.24$ day and
$\beta =0.90\pm 0.12$.  The parameters are strongly correlated, with a
normalized correlation coefficient $\rho(\alpha,\beta) =-0.99$, and
$\chi^2 = 25.94$, which approaches a low probability for 18 degrees of
freedom ($\chi^2/{\rm dof} = 1.44$ and $P(\chi^2|{\rm dof}) = 0.05$
for a one-tailed $\chi^2$ test).  Since there is good reason to
suspect that the \emph{u {\rm and} U} bands are affected by BLR
emission (see \S4), we also fit the data excluding these lags.  With
these bands excluded, we find $\alpha =$\ \alphafit\ day and $\beta
=$\ \betafit.  The normalized correlation coefficient does not change,
but the goodness-of-fit is now $\chi^2 = 16.85$ with ${\rm dof} = 16$,
and $\chi^2/{\rm dof} = 1.05$ (and $P(\chi^2|{\rm dof}) = 0.60$ for
the same one-tailed test).  The interpretation of Equation
\ref{equ:lag_wv} is discussed in \S5.3.

\ifapj
  \begin{deluxetable*}{rrrrrrr}
\else
  \begin{deluxetable}{rrrrrrr}
\fi
\tablewidth{0pt}
\tablecaption{Time Delays \label{tab:lags}}
\tablehead{& & & \colhead{ICCF}& & & \colhead{JAVELIN}\\
\colhead{Source} & \colhead{Filter} & \colhead{$\lambda_{\rm pivot}$} & \colhead{$\tau_{\rm cent}$} & \colhead{$\tau_{\rm peak}$} & \colhead{$r_{\rm max}$} & \colhead{$\tau_{\rm JAV}$}\\
& &\colhead{(\AA)}& \colhead{(days)} & \colhead{(days)} & &\colhead{(days)}
}
\startdata
\emph{Swift} & HX& 4.4 & $-0.65_{-0.45}^{+0.45}$ & $-0.46 _{- 0.39}^{+ 0.49}$ &$ 0.35\pm0.20$ & \dots\\
\emph{Swift} & SX& 25.3 & $0.08 _{- 0.51}^{+ 0.51}$ & $0.23 _{- 0.39}^{+ 0.29}$ & $ 0.44\pm 0.07$ & \dots\\
\emph{HST} & $\lambda 1158$ & 1158 & $-0.17 _{- 0.16}^{+ 0.16}$ & $-0.21 _{- 0.10}^{+ 0.08}$ & $ 1.07 \pm  2.53$ & $-0.14 _{- 0.04}^{+ 0.04}$ \\
\emph{HST} & $\lambda 1479$ & 1479 & $ 0.15 _{- 0.16}^{+ 0.18}$ & $ 0.14 _{- 0.06}^{+ 0.23}$ & $ 1.03 \pm  1.08$ & $ 0.03 _{- 0.04}^{+ 0.04}$ \\
\emph{HST} & $\lambda 1746$ & 1746 & $ 0.22 _{- 0.19}^{+ 0.16}$ & $ 0.21 _{- 0.10}^{+ 0.10}$ & $ 0.98 \pm  0.01$ & $ 0.14 _{- 0.05}^{+ 0.05}$ \\
\emph{Swift} & {\it UVW2} & 1928 & $ 0.63 _{- 0.18}^{+ 0.19}$ & $ 0.59 _{- 0.10}^{+ 0.20}$ & $ 0.92 \pm  0.16$ & $ 0.68 _{- 0.09}^{+ 0.11}$ \\
\emph{Swift} & {\it UVM2} & 2246 & $ 0.68 _{- 0.20}^{+ 0.19}$ & $ 0.59 _{- 0.10}^{+ 0.20}$ & $ 0.90 \pm  0.27$ & $ 0.69 _{- 0.17}^{+ 0.14}$ \\
\emph{Swift} & {\it UVW1} & 2600 & $ 0.93 _{- 0.23}^{+ 0.20}$ & $ 0.88 _{- 0.20}^{+ 0.29}$ & $ 0.89 \pm  0.01$ & $ 0.90 _{- 0.16}^{+ 0.16}$ \\
\emph{Swift} & {\it U} & 3467 & $ 1.80 _{- 0.24}^{+ 0.24}$ & $ 1.47 _{- 0.29}^{+ 0.20}$ & $ 0.88 \pm  0.35$ & $ 1.62 _{- 0.16}^{+ 0.16}$ \\
Ground & {\it u} & 3472 & $ 2.03 _{- 0.39}^{+ 0.43}$ & $ 2.04 _{- 0.39}^{+ 0.29}$ & $ 0.83 \pm  0.04$ & $ 1.90 _{- 0.04}^{+ 0.04}$ \\
Ground & {\it B} & 4369 & $ 1.42 _{- 0.33}^{+ 0.36}$ & $ 1.22 _{- 0.29}^{+ 0.20}$ & $ 0.91 \pm  0.02$ & $ 1.36 _{- 0.13}^{+ 0.11}$ \\
\emph{Swift} & {\it B} & 4392 & $ 1.64 _{- 0.27}^{+ 0.31}$ & $ 1.28 _{- 0.39}^{+ 0.29}$ & $ 0.82 \pm  0.02$ & $ 1.34 _{- 0.21}^{+ 0.19}$ \\
Ground & {\it g} & 4776 & $ 1.98 _{- 0.29}^{+ 0.34}$ & $ 1.64 _{- 0.39}^{+ 0.29}$ & $ 0.89 \pm  0.02$ & $ 1.45 _{- 0.04}^{+ 0.06}$ \\
Ground & {\it V} & 5404 & $ 2.04 _{- 0.20}^{+ 0.22}$ & $ 1.87 _{- 0.10}^{+ 0.29}$ & $ 0.84 \pm  0.02$ & $ 1.72 _{- 0.07}^{+ 0.07}$ \\
Ground & {\it r} & 6176 & $ 3.13 _{- 0.46}^{+ 0.41}$ & $ 3.12 _{- 0.59}^{+ 0.29}$ & $ 0.85 \pm  0.04$ & $ 2.38 _{- 0.07}^{+ 0.06}$ \\
Ground & {\it R} & 6440 & $ 3.22 _{- 0.29}^{+ 0.30}$ & $ 2.88 _{- 0.20}^{+ 0.39}$ & $ 0.87 \pm  0.11$ & $ 2.81 _{- 0.05}^{+ 0.04}$ \\
Ground & {\it i} & 7648 & $ 3.99 _{- 0.29}^{+ 0.29}$ & $ 3.90 _{- 0.29}^{+ 0.20}$ & $ 0.90 \pm  0.02$ & $ 3.46 _{- 0.08}^{+ 0.11}$ \\
Ground & {\it I} & 8561 & $ 3.59 _{- 0.54}^{+ 0.53}$ & $ 2.88 _{- 0.88}^{+ 0.59}$ & $ 0.86 \pm  0.06$ & $ 3.38 _{- 0.07}^{+ 0.07}$ \\
Ground & {\it z} & 9157 & $ 3.93 _{- 0.40}^{+ 0.44}$ & $ 3.71 _{- 0.20}^{+ 0.59}$ & $ 0.84 \pm  0.04$ & $ 3.88 _{- 0.06}^{+ 0.08}$ \\
\enddata
\tablecomments{Measured relative to the \emph{HST} \drivelc\ light
  curve and corrected to the rest frame. The \emph{Swift} lags are
  recalculated from Paper II using a second-order polynomial
  detrending routine, as described in \S3.}

 \ifapj
  \end{deluxetable*}
\else
  \end{deluxetable}
\fi

\ifapj
  \begin{deluxetable*}{lrrrrr}
\else
  \begin{deluxetable}{lrrrrr}
\fi
\tablecaption{Parameters for lag--wavelength fits\label{tab:fitparams}}
\tablehead{\colhead{Model} & \colhead{$\alpha$ (days)} & \colhead{$\beta$} & \colhead{${\rho(\alpha,\beta)}$} & \colhead{$\chi^2$} & \colhead{$\chi^2/dof$}}
\startdata
All & $0.97 \pm 0.24$ & $0.90 \pm 0.12$ & -0.99 & 25.94 & 1.44 \\
 & $0.43 \pm 0.02$ & 4/3 &  & 38.66 & 2.03 \\
No {\it Uu} & $0.79 \pm 0.22$ & $0.99 \pm 0.14$ & -0.99 & 16.85 & 1.05 \\
 & $0.42 \pm 0.02$ & 4/3 &  & 22.64 & 1.33 \\
No {\it UuIR} & $0.58 \pm 0.20$ & $1.18 \pm 0.19$ & -0.99 & 12.4 & 0.89 \\
 & $0.45 \pm 0.02$ & 4/3 &  & 13.0 & 0.87 \\
\enddata
\ifapj
  \end{deluxetable*}
\else
  \end{deluxetable}
\fi

\section{ Contamination by Broad-Line Region Emission}
As noted above, the \emph{u {\rm and} U} lags are outliers from the
trend in Figure \ref{fig:lag_wv}. A major component of the flux
observed in these filters is the ``small blue bump,'' caused by
bound-bound and bound-free hydrogen emission (the so-called Balmer
continuum), as well as blended Fe\,{\sc ii} lines that originate in
the BLR.  This BLR emission may cause the \emph{u} and \emph{U}-band
lags to be biased estimators of the light-crossing time within the
continuum source.  In fact, several filters pick up other spectral
features that originate in the BLR.  The strongest is the prominent
H$\alpha$ line in the \emph{r} and \emph{R} bands, although additional
emission lines and a diffuse continuum consisting of bound-free,
free-free, electron scattering, and reflection is expected to be
present at all wavelengths \citep{Korista2001}.  Understanding the
impact of BLR emission on the observed lags is therefore important for
interpreting the interband time delays.

In this section, we assess the effect of BLR emission on the
interband continuum lags.  First, we decompose spectra of NGC 5548
into models of each emission component.  We then estimate the
fractional contribution from BLR emission in each filter using
synthetic photometry.  Finally, we simulate broad-band filter
observations by combining mock continuum and BLR light curves, and
search for biases in the lags by cross-correlating each emission
component with the \drivelc\ light curve.

\subsection{Spectral Decomposition}

We begin by decomposing spectra of NGC 5548 into models of each
emission component.  We obtained moderate-resolution ($R \approx 2000$)
optical spectra of NGC 5548 using the Multi-Object Double
Spectrographs \citep[MODS; ][]{Pogge2010} on the Large Binocular
Telescope (LBT; \citealt{Hill2010}).  These observations are from 2014
June 08 and 2014 June 25 UT (HJD =2,456,817 and 2,456,834, respectively).
The spectra were reduced and flux-calibrated using the {\tt modsIDL}
Spectral Reduction Pipeline.\footnote{A full description can be found
  at \url{
    http://www.astronomy.ohio-state.edu/MODS/Manuals/modsIDL.pdf}} The
spectra cover the wavelength range from 3000\,\AA\ to 1 $\mu$m.
Wavelength solutions were derived from comparison-lamp calibrations
for each observing run.  Relative-flux calibration was performed using
three standard stars observed on the same nights as NGC 5548; however, the
observations were taken in poor atmospheric conditions, making their
absolute-flux calibration unreliable.  We therefore rescaled the
spectra so that the integrated [O\,{\sc iii}]\,$\lambda 5007$ fluxes
match the value measured for the photometric nights of the optical
spectral RM campaign, $(5.01\pm 0.11) \times 10^{-13} \,\text{erg\
  s}^{-1}\text{ cm}^2$ (\citePei).  The slit width and extraction
window of the MODS spectra were 5\arcsec\ and 15\arcsec, respectively,
chosen to match those of the optical monitoring spectra.  This ensures
that the relative contribution of host-galaxy light, narrow-line
emission, and BLR emission are the same in both datasets.  We
corrected for Galactic extinction following the prescription described
in \S2.4.  We did not make any correction for telluric absorption
because broad-band filters suffer from the same effect.

Since we are only concerned with the relative magnitude of various
emission components to the broad-band filter fluxes, we employed a
minimal spectral decomposition, which is relatively coarse compared to
state-of-the-art spectral modeling.  Accordingly, we do not interpret
any of our model parameters as indicative of physical conditions
within the AGN, and instead focus on finding a model that provides a
good fit to the data (based on minimizing $\chi^2$).  Our
decomposition has three components: host-galaxy starlight, the
underlying AGN continuum, and the Balmer continuum shortward of
$\sim\! 3648$\,\AA\ (rest frame).  We ignore the diffuse continuum at
other wavelengths, since it is poorly constrained, while the Balmer
continuum can be determined from the shape and amplitude of the small
blue bump.  Emission-line fluxes are then estimated by subtracting the
summed model components from the observed spectrum.

We simultaneously fit each component with an MCMC calculation, masking
AGN emission lines and telluric absorption.  We also masked the long
and short edges of the spectra, because the MODS flux calibration is
unreliable at $\lambda < 3200$\,\AA\ and $\lambda > 9100$\,\AA\ (rest
frame).  At these wavelengths, we set the observed flux equal to the
summed model, which implicitly sets the emission-line flux to zero.
This has a small effect on the estimated BLR contamination in the
\emph{u}, \emph{U}, \emph{I}, and \emph{z} bands, but is more robust
than using the unreliable flux calibration.

Details of the model components are as follows.
\begin{enumerate}
\item \textbf{Host Galaxy:} We determined the host-galaxy spectrum
  using the {\tt STARLIGHT} spectral synthesis code
  \citep{Fernandes2004}.  {\tt STARLIGHT} fits the observed spectrum
  with a linear combination of a large library of synthetic stellar
  populations that span a wide range of ages and metallicities (150
  templates from \citealt{Bruzual2003}).  The best-fitting models
  consist of several very old (usually $>\!10^{10} $ year) stellar
  populations at a range of metallicities ($0.4$--$2.5\, {\rm Z}_{\odot}$),
  and provide a reasonable match to the galaxy templates used by
  \citet{Denney2010} and \citet{Mehdipour2015}.  The resulting host
  templates have one parameter, the flux normalization.  We also
  impose a tight prior on the flux at 5100\,\AA\ (rest frame), chosen
  to match the value measured by \citet{Bentz2013} adjusted to the
  MODS slit width and extraction window, ($4.52 \pm 0.45) \times
  10^{-15}\,\text{erg\ s}^{-1}\text{ cm}^2\text{ \AA}^{-1}$.

\item \textbf{Power Law:} A broken power law is used to model the AGN
  continuum emission.  This component has four free parameters: a
  flux-normalization factor, two spectral indices, and the location of
  the transition between indices.  A loose prior (a Gaussian
  distribution with mean 5700\,\AA\ and width 700\,\AA) is imposed on
  the transition wavelength, to prevent it from moving to the edges of
  the spectra.

\item \textbf{Balmer Continuum:} The Balmer continuum component is
  estimated from a grid of models calculated by \cite{Dietrich2002},
  evaluated at varying temperatures, electron densities, and optical
  depths.  Again, we simply choose the template that produces the
  overall minimum value of $\chi^2$.  The templates have a single
  parameter, a flux rescaling factor.
\end{enumerate}

We ignored blended Fe\,{\sc ii} emission, because Fe emission is
relatively weak in NGC 5548 \citep{Denney2009,Mehdipour2015} and
varies with an amplitude $<$50--75\% that of H$\beta$
\citep{Vestergaard2005}. This component is therefore expected to
contribute very little flux to the broad-band photometric measurements
and have a negligible impact on the observed lag.  In order to assess
the effect of this omission, we also fit the spectra with the small
blue bump template of \cite{Mehdipour2015}, which includes blended
Fe\,{\sc ii} emission lines.  We found that these templates produce a
poorer fit than the \citet{Dietrich2002} templates at the blue end of
the spectrum, which may be a result of the limited wavelength coverage
of our MODS spectra in the near-UV.

\begin{figure*}
\includegraphics[width=1.0\textwidth]{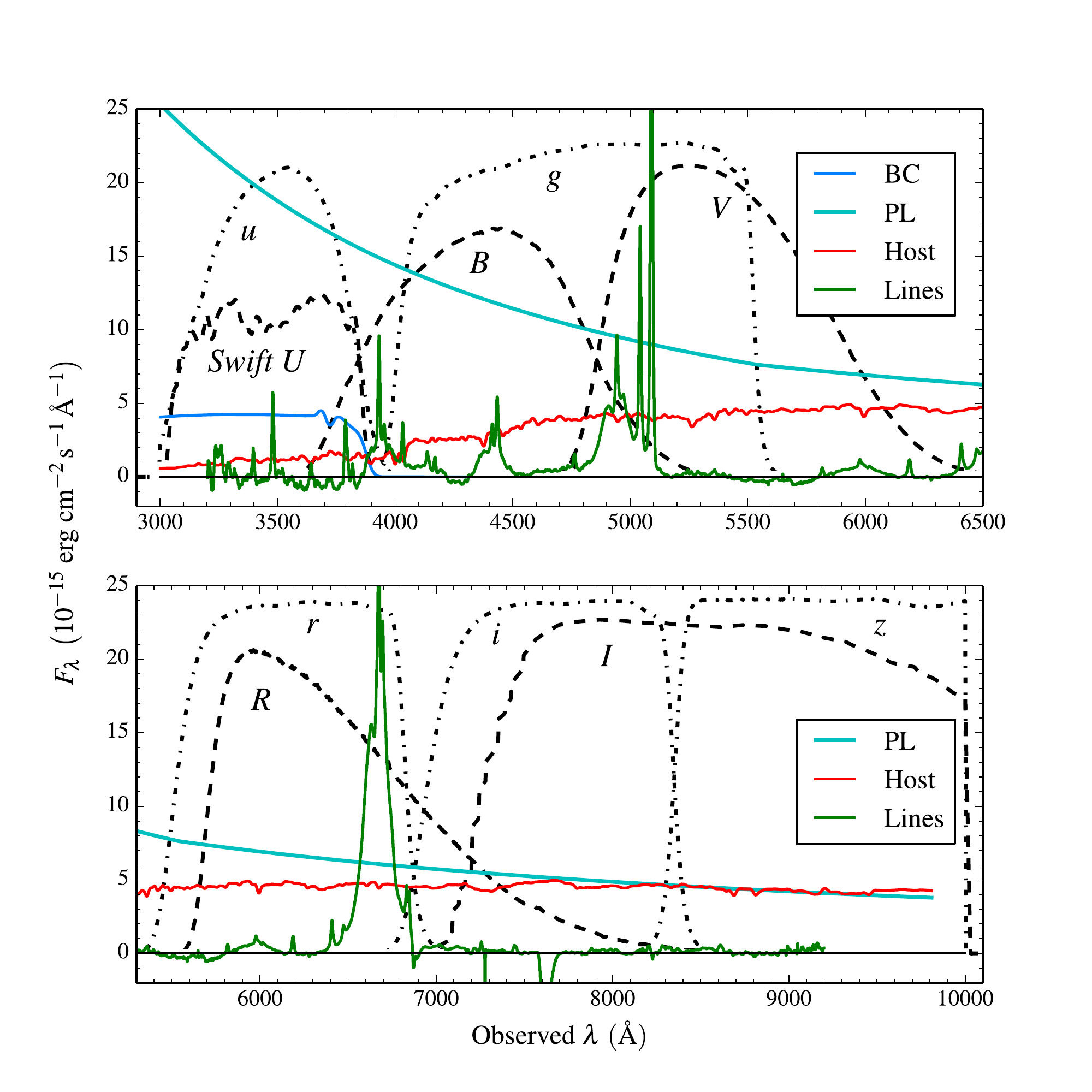}
\caption{Decompositions of the MODS spectra from 2014 June 08, showing
  the contribution of the model components to different filters.
  ``BC'' is the Balmer continuum, ``PL'' is the power law, ``Host'' is
  the host-galaxy component, and ``Lines'' are the AGN emission lines.
  The emission lines are estimated by subtracting the total model from
  the observed spectrum.  Johnson/Cousins optical filter transmission
  curves (and {\it Swift U}) are shown by the dashed black lines, SDSS
  filters are shown by the dot-dashed lines.  The {\it Swift U} and
  {u} bands are truncated at 3000\,\AA\ and the \emph{I} and \emph{z}
  bands are truncated at 1 $\mu$m, in order to represent the
  atmospheric transmission cutoff.\label{fig:filters} }
\end{figure*}

Each epoch was fit independently, and the resulting component
parameters are in reasonable agreement, after allowing for the
intrinsic variability of the power-law and Balmer continuum.  The flux
rescaling factors of the power-law continuum and galaxy templates are
degenerate, so the prior imposed on the host-galaxy flux at 5100\,\AA\
(rest frame) does the most to constrain these parameters.  Figure
\ref{fig:filters} shows an example of the decomposition, using the
spectrum from 2014 June 08, overlaid with the filter transmission
curves.

\subsection{Synthetic Photometry}

Next, we estimate the contribution of each model component to the
observed flux in each broad-band filter.  We first reapply Galactic
reddening to the model components, since differential extinction may
affect the integrated flux across broad-band filters.  We then
calculate the observed flux using the {\tt synphot} {\tt IRAF} task
and filter transmission curves for the calibration telescopes (WC18
\emph{BVRI} filters and LT \emph{ugriz} filters), truncated at
3000\,\AA\ and 1$\mu$m to represent the atmospheric transmission
cutoff.  Uncertainties in the broad-band fluxes of individual
components were estimated by resampling the posterior distributions of
the model component parameters and rerunning {\tt synphot} $10^3$
times.

Table \ref{tab:fracs} shows the results of our synthetic photometry.
The ``Total'' column was calculated from the original spectrum, and
the fractional contributions of individual components are reported
relative to this value.  The uncertainties represent the central 68\%
confidence interval of the resampled synthetic photometry
distributions.  The uncertainties are generally less than 1\% because
of the tight prior on the 5100\,\AA\ host-galaxy flux, which forces
the galaxy template to be nearly constant and limits the variation of
the other model components.

We do not consider effects of changing detector sensitivity with
wavelength, since quantum efficiency curves for different instruments
are usually much more variable than their filter transmission curves.
Quantum efficiency will have the largest impact on the {\it I} and
{\it z} filters, limiting the response of these filters at wavelength
shorter than the cutoff imposed at 1\,$\mu$m.  We investigated this
effect by truncating the filter response at 9000\,\AA\ and repeating
the experiment (essentially simulating a very steep quantum efficiency
curve).  We found that the final fractional contributions of the
host/power-law components in these bands changes by 1\% or less, and
is therefore of minimal importance for our conclusions.

We find that the power-law component is dominant from the \emph{u}
band through the \emph{V} band ($>50$\% of the flux), although the
host galaxy makes considerable contributions even in the \emph{B} band
($\sim \!20$\%).  At longer wavelengths, the power-law component and
host galaxy contribute roughly equal amounts of flux, except for the
\emph{r} and \emph{R} bands, which include a substantial contribution
from the H$\alpha$ line: 20\% in the \emph{r} band and 15\% in the
\emph{R} band.  Line emission in all other filters is $\leq\! 10$\%.
Balmer continuum emission accounts for about 19\% of the flux in the
\emph{u} and \emph{U} filters. The \cite{Mehdipour2015} blended Fe
templates contribute $<1$\% of the observed flux in the \emph{g},
\emph{V}, and \emph{r} bands, confirming that Fe emission is a
negligible component of the broad-band fluxes of this object.

\subsection{Impact on Time Delays}
The final step is to estimate the impact of BLR emission on the
recovered interband time delays.  First, we simulate light curves for
the AGN continuum, Balmer continuum, and BLR emission models.  We then
sum the component light curves to reproduce light curves as would be
observed in a given filter, and calculate the lag between the
composite light curve and the \emph{HST} \drivelc\ continuum light
curve.

The observed light curve is a superposition of the continuum emission
and BLR emission,
\begin{align}
\label{equ:model}
X_{\rm obs}(t) = c(t) + l(t),
\end{align}
where $X_{\rm obs}$ is the observed light curve in filter $X$, $c(t)$
is the continuum light curve in that filter, and $l(t)$ is the line
light curve, assumed to originate in the BLR.  We use the term ``line
light curve'' to refer to any emission produced in the BLR, including
the Balmer continuum.

To simulate $c(t)$, we calculated the lag $\tau_{\rm cont}$ implied by
the best-fit parameters in Figure \ref{fig:lag_wv} ($\alpha
=$\ \alphafit\ and $\beta = $\ \betafit\ in Equation \ref{equ:lag_wv}) at
the pivot wavelength of the filter, and shifted the {\tt JAVELIN} DRW
model of the \emph{HST} \drivelc\ light curve by this amount. This
method assumes that the \emph{HST} \drivelc\ light curve drives $c(t)$
through instantaneous reprocessing after some light-travel-time delay,
as would be expected for X-ray reprocessing in the accretion disk.
\footnote{ Reprocessed emission is also expected to be somewhat
  smoothed in time compared to the driving light curve.  We therefore
  also considered versions of $c(t)$ which are both smoothed and
  shifted by convolving the {\tt JAVELIN} \drivelc\ model with a
  top-hat function of amplitude $1/(2\tau_{\rm cont})$ for $0<\tau <
  2\tau_{\rm cont}$.  We found that this smoothing made very little
  difference on the results, and so we only discuss the results for
  the shifted versions of $c(t)$ here.}

In RM, the line emission is assumed to be powered by ionizing
continuum emission, so that
\begin{align}
l(t) = \int \Psi(\tau)C(t-\tau)\, d\tau,
\end{align}
where $C(t)$ is the driving continuum light curve and
$\Psi(\tau)$ is the transfer function.  For simplicity, we assume
$C(t)$ equal to the {\tt JAVELIN} model of the \emph{HST} \drivelc\
light curve and a top-hat transfer function,
\begin{align}
  \Psi(\tau) = \frac{1}{w} \mbox{ for }(\bar \tau - w/2) < \tau <
  (\bar \tau + w/2),
\end{align}
where $\bar \tau$ is the mean line lag and $w$ is the width of the
smoothing kernel.  The choice of a top-hat function is for
mathematical convenience and does not reflect any particular
geometry, although it is widely consistent with a range of BLR
configurations (for example, a spherical shell or the gross properties
of an inclined disk/annulus; \citealt{Peterson2001}). We varied $\bar
\tau$ and $w$ by octaves, with $\bar \tau$ = 2, 4, 8, and 16 days, and
$w$ = 0, 2, 4, 8, and 16 days.  These values were chosen to sample the
parameter space near the mean H$\beta$ lag during the monitoring
campaign ($8.57\pm 0.67$ days; \citePei).  To a low approximation, the
Balmer continuum and H$\alpha$ lag would be expected to lie near this
value.  Finally, we enforced causality by setting $\Psi(\tau) = 0$ for
$\tau < 0$.

\begin{figure*}
  \includegraphics[width=\textwidth]{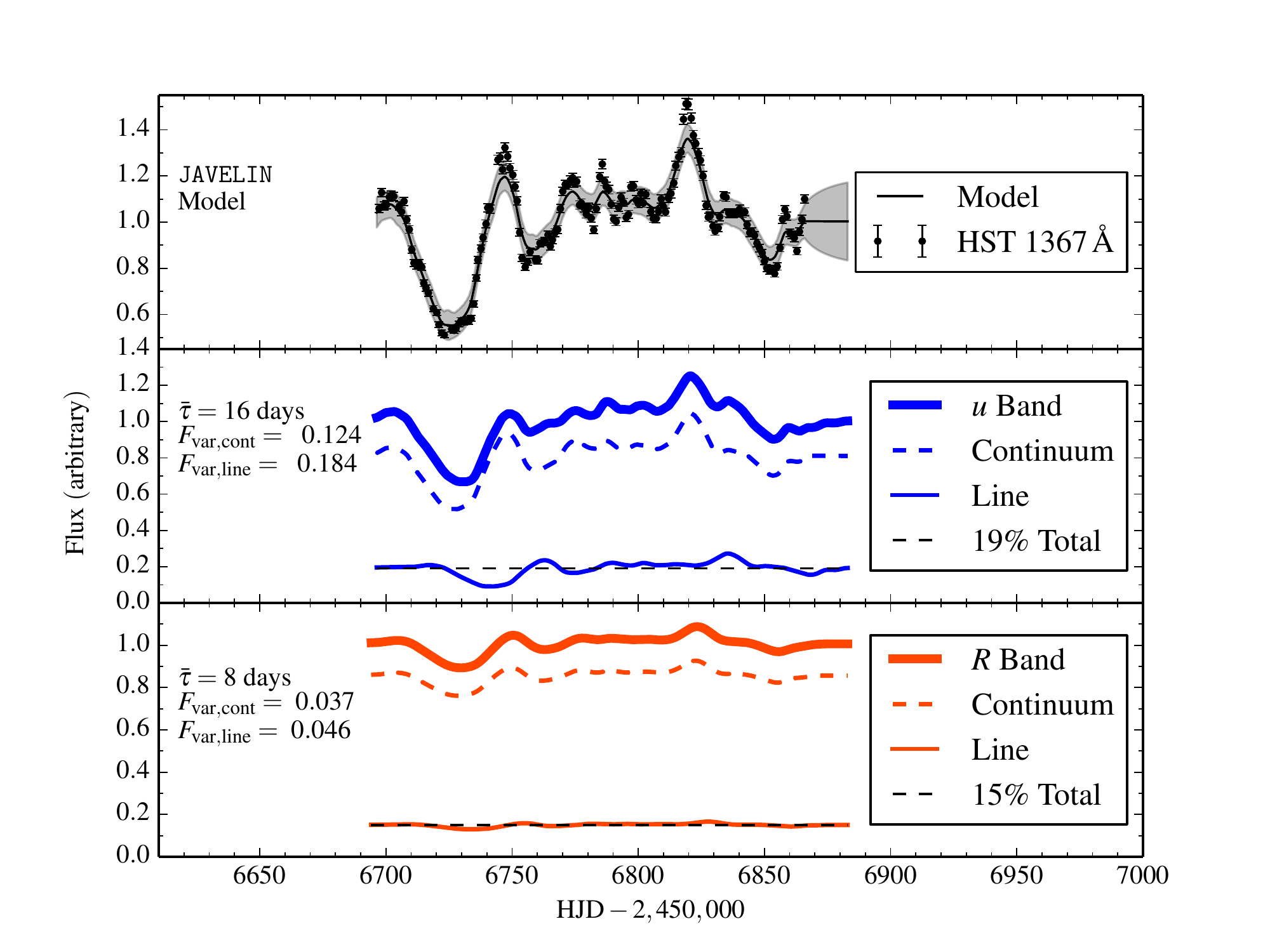}
  \caption{Examples of mock light curves, $c(t)$, $l(t)$, and $X_{\rm obs}
    = c(t) + l(t)$, used for the analysis in \S4.3.  The top panel
    shows the \emph{HST} \drivelc\ light curve and the {\tt JAVELIN}
    model used to generate the mock light curves, with the 1$\sigma$
    uncertainty shown by the grey band.  The middle panel displays an
    example of a mock \emph{u}-band light curve, with a large line lag
    and high fractional variability, likely to result in the largest
    change of the observed lag.  The bottom panel shows an example of
    a mock \emph{R}-band light curve, with a more realistic line lag
    and fractional variability, chosen to be consistent with the
    H$\beta$ light curve. See \S4.3 for further
    details. \label{fig:mock_lc} }
\end{figure*}

We simulate light curves for the \emph{u, U, r, {\rm and} R} bands, in
order to investigate the impact of Balmer continuum and H$\alpha$
emission on the recovered lags.  After generating the grid of shifted
and smoothed line light curves, we renormalized each so as to
reproduce the level of BLR contamination inferred from the spectral
decomposition (Table \ref{tab:fracs}).  We then adjusted the
fractional variability amplitude $F_{\rm var}$ (defined in \S2.4) of
both the continuum and line light curves to match their observed
values.  For the continuum light curves, $F_{\rm var,cont}$ is
estimated directly from the observed broad-band light curves (Table
\ref{tab:lcprop}, Column 9).  For the line light curves, we set $F_{\rm
  var,line} = 4.6\%$, derived from the observed H$\beta$ light curve
(\citePei).  We also experimented with changing the fractional
variability amplitude of the line light curve to $F_{\rm var,line} = $
0.012, 0.023, 0.092, and 0.184.  Examples of two composite light curves
and their model components, $c(t)$ and $l(t)$, are shown in Figure
\ref{fig:mock_lc}.

\begin{figure*}
  \includegraphics[width=\textwidth]{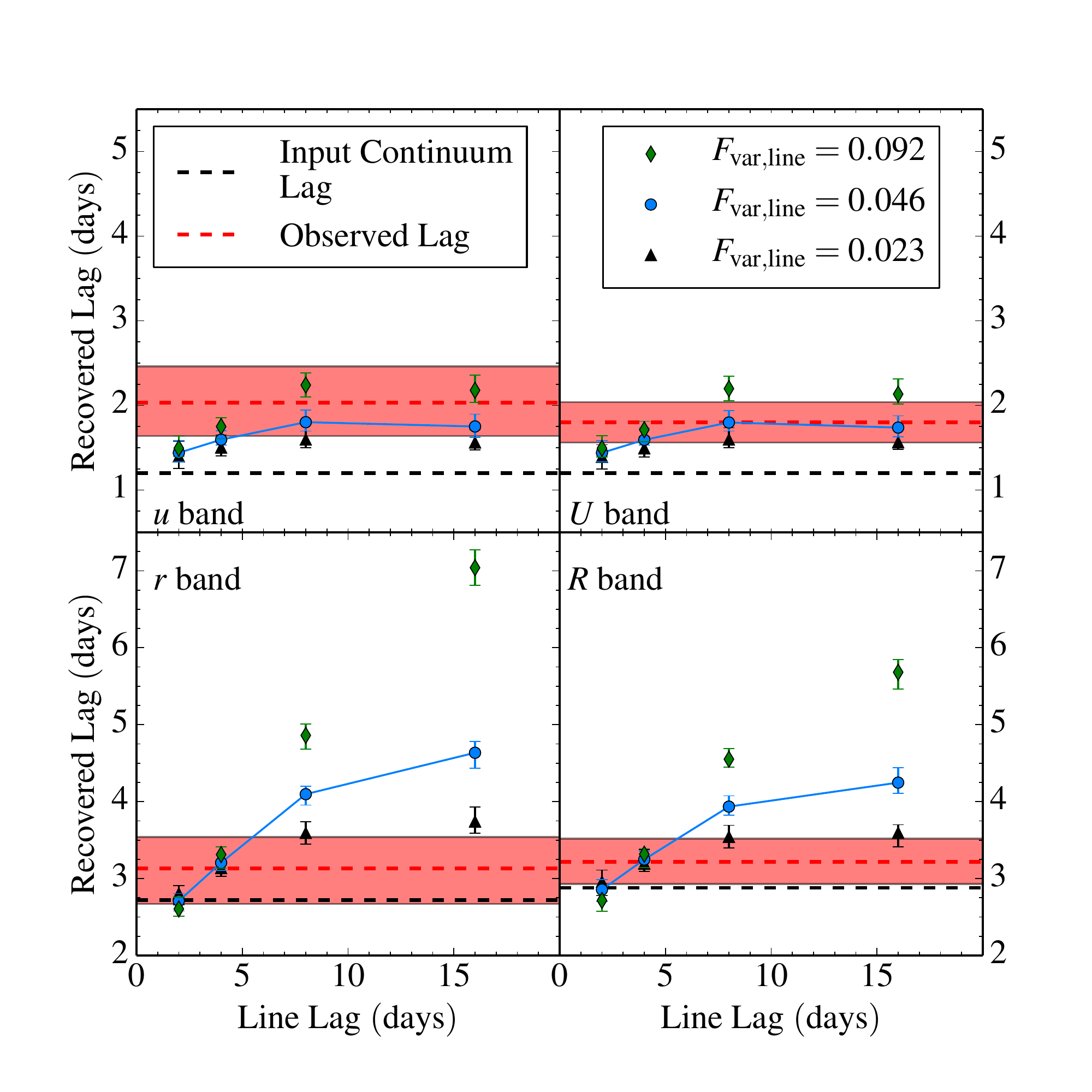}
  \caption{Recovered lags of mock light curves as a function of input
    line light curve $l(t)$ lag.  The colored points show the results
    for different variability amplitudes of the line light curve.  The
    solid blue lines indicate the variability amplitude observed in
    the H$\beta$ light curve.  The black dashed line represents the
    input lag of the continuum light curve $c(t)$, while the red
    dashed line is the observed lag and the red band is its 1$\sigma$
    uncertainty.  See \S4.3 for further details.  \label{fig:sim_lag1}
  }
\end{figure*}

After constructing $c(t)$ and $l(t)$ for each model, we calculated the
lags of these light curves relative to $C(t)$ using the ICCF method
described in \S3.  In all cases, we recovered the input values
of $\bar \tau$ and $\tau_{\rm cont}$ to within the time resolution of
the model light curves (0.12 day).  We then calculated the ICCF for
the composite light curve $c(t) + l(t)$, finding that the recovered
lags are most sensitive to the choice of $F_{\rm var,line}$ and $\bar
\tau$ but are virtually independent of $w$.  The resulting mean lags
are shown in Figure \ref{fig:sim_lag1} as a function of input $\bar
\tau$ for the three values of $F_{\rm var,line}$ near that of H$\beta$
(larger or smaller values of $F_{\rm var,line}$ do not plausibly reproduce
the observed lags, and are omitted for clarity).  Larger values of
these parameters tend to increase the recovered lag, but at the
fiducial values of H$\beta$ the change is 0.6--1.2 days (blue point in
Figure \ref{fig:sim_lag1} with $\bar \tau = 8 $).  

We also checked for an effect of BLR contamination on the lag
uncertainties.  For each model, we found that larger values of $F_{\rm
  var,line}$ and $\bar \tau$ tend to increase the width of the ICCF.
However, there was no correlation between these parameters and the
location or width of the ICCF centroid distribution.  This means that
the lag uncertainties depend more sensitively on the light-curve
quality rather than on any BLR contamination.

For values of $F_{\rm var,line} $ that are smaller than $F_{\rm
  var,cont}$ ($F_{\rm var,line}\leq 0.023$ in {\it r} and {\it R} and
$\leq 0.092$ in {\it u} and {\it U}), the input line lag only has a
limited effect on the recovered lag, evidenced by the flattening of
the trends in Figure \ref{fig:sim_lag1}.  This result is in contrast
to the simple expectation that the observed lag is the flux-weighted
mean lag of the line and continuum light curves, which scales linearly
with the line lag.  Instead, it appears that the observed lag only
follows the line lag if the BLR emission dominates the variability
properties of the composite light curve, as seen for the mock {\it
  r} and {\it R} bands at large $F_{\rm var,line}$.  This indicates that
the bias of the continuum lag introduced by BLR emission will usually
be limited for broad-band filter light curves that are dominated by
continuum emission, although the bias may still be important for small
continuum lags.

Our simulations with these fiducial H$\beta$ parameters produce
\emph{u} and {\it U}-band lags in excellent agreement with the
observed lags, while the simulated {\it r} and {\it R}-band lags
overestimate the observed lag by about 1 day.  Our current campaign
cannot directly address the issue of the unknown values of $F_{\rm
  var,line}$ and $\bar \tau$ for these reverberations.  However, it is
expected from photo-ionization modeling that the Balmer continuum has
a larger response ($F_{\rm var,line}$) but shorter lag than H$\beta$, while
H$\alpha$ should have a smaller response but longer lag
\citep{Korista2001,Korista2004}.  Based on Figure \ref{fig:sim_lag1},
this would serve to reduce the discrepancy between the recovered and
observed lag in the {\it r} and {\it R} bands, while the recovered and
observed lag in the {\it u} and {\it U} bands would remain in good
agreement.

Thus, our simulations suggest that contamination by BLR emission can
reasonably account for the systematic offset of the measured \emph{u,
  U, r, {\rm and} R} lags above the fit in Figure \ref{fig:lag_wv}.
This bias is well resolved in the {\it u} and {\it U} bands (the
offset from the fit in Figure \ref{fig:lag_wv} is $2.0\sigma$ and
$2.5\sigma$, respectively), but of small importance in the {\it r} and
{\it R} bands ($0.6\sigma$ and $1.7\sigma$, respectively).  This result
justifies our exclusion of the {\it u} and {\it U}-band data in the
fit to Equation \ref{equ:lag_wv}.

\citet{Chelouche2013a} and \citet{Chelouche2013b} claim that BLR
emission is responsible for the large {\it B}-{\it R}/Cousins {\it I}
lags in the \citet{Sergeev2005} NGC 5548 light curves, and they
find optical continuum lags consistent with 0 days.  This is at odds
with our results, since the BLR biases would have to be $\gtrsim\! 8$\
days.  These studies use a variation of the ICCF method (the
multivariate CCF) to disentangle line and continuum lags from
emission observed in a single filter.  As we have already noted, gaps
in the \citet{Sergeev2005} data make cross-correlation functions that
rely on interpolation unreliable.  Furthermore, this bias would imply
that line emission contributes 30--50\% of the flux in the {\it R} and
Cousins {\it I} bands, which is implausibly high based on both our
spectral decompositions and the composite Seyfert 1 spectrum of
\citet{Chelouche2013b}.  

\ifapj
  \begin{deluxetable*}{lcrrrr}
\else
  \begin{deluxetable}{lrrrrr}
\fi
\tablewidth{0pt}
\tablecaption{Flux percentage contribution by spectral component.\label{tab:fracs}}
\tablehead{\colhead{Filter} & \colhead{Total} & \colhead{PL} & \colhead{BC} & \colhead{Host} & \colhead{Lines}\\
&\colhead{($\rm 10^{-11} erg\,\,cm^{-2}\,s^{-1}$)}&\colhead{(\%)}&\colhead{(\%)}&\colhead{(\%)}&\colhead{(\%)}
}
\startdata

2014 June 08\\
\hline
\emph{U} & 8.42 & $  76.8 \pm  1.5 $ & $  16.7 \pm  0.5 $ & $  4.9 \pm  0.1 $ & $  2.3 \pm  1.6 $ \\
\emph{u} & 8.43 & $  76.6 \pm  1.5 $ & $  16.7 \pm  0.5 $ & $  4.8 \pm  0.1 $ & $  2.7 \pm  1.6 $ \\
\emph{B} & 7.23 & $  72.6 \pm  0.9 $ & $  1.4 \pm  0.0 $ & $  18.2 \pm  0.5 $ & $  7.6 \pm  1.0 $ \\
\emph{g} & 7.40 & $  65.6 \pm  0.6 $ & \dots & $  24.3 \pm  0.7 $ & $  10.0 \pm  0.8 $ \\
\emph{V} & 7.39 & $  59.0 \pm  0.5 $ & \dots & $  32.1 \pm  0.9 $ & $  9.1 \pm  1.0 $ \\
\emph{r} & 8.91 & $  47.1 \pm  0.4 $ & \dots & $  33.4 \pm  0.9 $ & $  19.6 \pm  1.0 $ \\
\emph{R} & 8.44 & $  48.9 \pm  0.5 $ & \dots & $  36.4 \pm  1.0 $ & $  14.8 \pm  1.1 $ \\
\emph{i} & 7.45 & $  53.5 \pm  0.7 $ & \dots & $  48.5 \pm  1.3 $ & \dots \\
\emph{I} & 5.77 & $  50.9 \pm  0.7 $ & \dots & $  50.3 \pm  1.2 $ & $  0.0 \pm  0.1 $ \\
\emph{z} & 4.60 & $  49.5 \pm  0.6 $ & \dots & $  50.7 \pm  1.0 $ & $  0.0 \pm  0.6 $ \\
\hline
2014 June 25\\
\hline
\emph{U} & 8.28 & $  72.3 \pm  1.5 $ & $  21.4 \pm  0.5 $ & $  5.1 \pm  0.1 $ & $  1.0 \pm  1.6 $ \\
\emph{u} & 8.29 & $  72.2 \pm  1.5 $ & $  21.4 \pm  0.5 $ & $  5.1 \pm  0.1 $ & $  0.6 \pm  1.4 $ \\
\emph{B} & 7.02 & $  69.7 \pm  1.0 $ & $  1.8 \pm  0.0 $ & $  19.9 \pm  0.2 $ & $  8.5 \pm  0.9 $ \\
\emph{g} & 7.21 & $  62.9 \pm  0.7 $ & \dots & $  26.6 \pm  0.3 $ & $  10.3 \pm  0.9 $ \\
\emph{V} & 7.29 & $  55.9 \pm  0.6 $ & \dots & $  34.9 \pm  0.4 $ & $  9.0 \pm  0.8 $ \\
\emph{r} & 8.93 & $  44.3 \pm  0.4 $ & \dots & $  35.8 \pm  0.4 $ & $  19.9 \pm  0.7 $ \\
\emph{R} & 8.47 & $  46.0 \pm  0.5 $ & \dots & $  39.0 \pm  0.4 $ & $  15.0 \pm  0.7 $ \\
\emph{i} & 7.53 & $  50.3 \pm  0.6 $ & \dots & $  51.9 \pm  0.5 $ & \dots \\
\emph{I} & 5.91 & $  47.8 \pm  0.5 $ & \dots & $  53.7 \pm  0.6 $ & \dots \\
\emph{z} & 4.73 & $  46.4 \pm  0.5 $ & \dots & $  54.0 \pm  0.5 $ & \dots \\
\hline
2014 June 08 \\(Blended Fe)\\
\hline
\emph{U} & 8.49 & $  82.6 \pm  0.9 $ & $  11.2 \pm  0.5 $ & $  5.3 \pm  0.1 $ & $  0.8 \pm  1.1 $ \\
\emph{u} & 8.49 & $  82.4 \pm  0.9 $ & $  11.7 \pm  0.5 $ & $  5.3 \pm  0.1 $ & $  0.9 \pm  1.1 $ \\
\emph{B} & 7.23 & $  73.5 \pm  0.8 $ & $  0.9 \pm  0.0 $ & $  20.0 \pm  0.3 $ & $  5.8 \pm  1.0 $ \\
\emph{g} & 7.40 & $  64.5 \pm  0.6 $ & $  0.4 \pm  0.0 $ & $  26.7 \pm  0.3 $ & $  8.4 \pm  0.8 $ \\
\emph{V} & 7.38 & $  55.9 \pm  0.5 $ & $  0.3 \pm  0.0 $ & $  35.3 \pm  0.5 $ & $  8.5 \pm  0.7 $ \\
\emph{r} & 8.90 & $  44.2 \pm  0.6 $ & $  0.1 \pm  0.0 $ & $  36.6 \pm  0.5 $ & $  19.2 \pm  0.6 $ \\
\emph{R} & 8.43 & $  45.7 \pm  0.6 $ & \dots & $  40.0 \pm  0.5 $ & $  14.4 \pm  0.7 $ \\
\emph{i} & 7.44 & $  49.3 \pm  0.9 $ & \dots & $  53.3 \pm  0.7 $ & \dots \\
\emph{I} & 5.77 & $  46.7 \pm  0.9 $ & \dots & $  55.0 \pm  0.7 $ & \dots \\
\emph{z} & 4.59 & $  45.1 \pm  0.9 $ & \dots & $  55.4 \pm  0.7 $ & \dots \\
\hline
2014 June 25 \\(Blended Fe)\\
\hline
\emph{U} & 8.41 & $  82.8 \pm  0.7 $ & $  11.4 \pm  0.4 $ & $  5.8 \pm  0.1 $ & \dots \\
\emph{u} & 8.39 & $  82.8 \pm  0.7 $ & $  12.0 \pm  0.5 $ & $  5.7 \pm  0.1 $ & \dots \\
\emph{B} & 7.02 & $  70.7 \pm  0.9 $ & $  0.9 \pm  0.0 $ & $  22.8 \pm  0.5 $ & $  5.7 \pm  1.2 $ \\
\emph{g} & 7.21 & $  60.7 \pm  0.8 $ & $  0.4 \pm  0.0 $ & $  30.5 \pm  0.7 $ & $  8.5 \pm  1.3 $ \\
\emph{V} & 7.27 & $  50.4 \pm  0.8 $ & $  0.3 \pm  0.0 $ & $  40.1 \pm  0.9 $ & $  9.4 \pm  1.2 $ \\
\emph{r} & 8.88 & $  38.4 \pm  0.7 $ & $  0.1 \pm  0.0 $ & $  41.3 \pm  0.9 $ & $  20.4 \pm  1.1 $ \\
\emph{R} & 8.40 & $  39.6 \pm  0.8 $ & \dots & $  45.2 \pm  1.0 $ & $  15.5 \pm  1.2 $ \\
\emph{i} & 7.46 & $  41.9 \pm  0.8 $ & \dots & $  60.3 \pm  1.3 $ & \dots \\
\emph{I} & 5.82 & $  38.9 \pm  0.9 $ & \dots & $  62.5 \pm  1.1 $ & \dots \\
\emph{z} & 4.65 & $  37.2 \pm  0.9 $ & \dots & $  63.2 \pm  0.9 $ & $  0.0 \pm  0.1 $ \\
\enddata
\tablecomments{PL is power law, BC is Balmer continuum, Host is the
  host galaxy, Lines are AGN emission lines.  BC includes a Fe
  emission template in the ``Blended Fe'' models.}
  \ifapj
  \end{deluxetable*}
\else
  \end{deluxetable}
\fi

\section{Discussion}

\subsection{UV/Optical Light Curves and Lags}
A primary goal of the AGN STORM project was to investigate how the
continuum emission changes as a function of wavelength, and to assess any
systematic issues introduced by using the optical continuum in place
of the far-UV or extreme-UV.  Figure \ref{fig:all_lc} shows a detailed
comparison of the \emph{HST} \drivelc\ light curve and all other data
used in this study.  We draw particular attention to the ground-based
\emph{V}-band light curve, since this is the most common choice of
ionizing continuum proxy in ground-based RM studies.  All of the major
events and salient characteristics of the \drivelc\ light curve are
reproduced in the \emph{V} band.  There are, however, several
noticeable differences.

\subsubsection{UV--Optical lags}
The first difference is a time delay between variations in the UV and
optical light curves.  Emission at 1158\,\AA, the shortest continuum
wavelength available in this study, probably originates from a region
of the accretion disk similar to that of the true ionizing continuum
at $\lambda \leq 912$\,\AA.  This is because the lag-wavelength
relation must flatten at small wavelengths (owing to the inner edge of
the disk), but the inner edge already makes an important contribution
to emission at $\sim\!  1000$\,\AA\ \citep{Novikov1973}. Extrapolating
the fit from Equation \ref{equ:lag_wv} to $\lambda = 912$\,\AA\
implies a 0.26 day lag relative to the \drivelc\ light curve, which is
in reasonable agreement with the \drivelc-1148\,\AA\ lag ($-0.16 \pm
0.16$ day).  We therefore adopt a value of 0.2 day for the lag
between the true ionizing continuum and the \drivelc\ emission, since
the lags for wavelengths $< 912$\,\AA\ are unlikely to be much
larger. This translates to a distance between the true ionizing
continuum and the optically emitting portion of the disk of $\sim 2.2$
light days.  A consequence of this UV-optical lag is that the radius
of the BLR in NGC 5548 is underestimated when derived from the
optical-H$\beta$ lag.  The optical--H$\beta$ lag is variable in time,
but typically has a value between 6 and 20 days
\citep{Peterson2004,Zu2011}.  Thus, if a similar UV--optical lag
exists in other AGN, the physical size of the BLR is being
systematically underestimated by up to $\sim\! 37\%$ (or $11\%$ for a
lag of 20 days).

This result does not affect current optical RM SMBH masses, because RM
only directly measures the virial product of the BLR, $c\tau (\Delta
V)^2/G$, where $\tau$ is the BLR lag and $\Delta V^2$ is its velocity
dispersion (estimated from line-profile widths). Since the geometry
and dynamics of the BLR are unknown, the virial product must be
rescaled by a factor $f$ in order to produce a SMBH mass.  While every
AGN has a different value of $f$, a statistical average $\langle f
\rangle$ can be calculated by calibrating an ensemble of virial
products to some other SMBH mass estimate.  Currently, this is done
using the $M-\sigma$ relation of local quiescent galaxies
\citep{Onken2004,Woo2010,Woo2013,Woo2015,Park12b,Park12a,Grier2013b}.
Thus, any systematic misestimation or bias of the lag (or velocity
dispersion) is compensated by the calibration of $\langle f\rangle $,
while the uncertainty of a single RM SMBH mass is dominated by the
statistical uncertainty in $\langle f\rangle$, currently about 
$25$--$33\%$ \citep{Grier2013b,Woo2015}.  However, any physical
interpretation of $\langle f \rangle$ (for example, a measure of the
mean inclination of the BLR, assuming a disk or otherwise flattened
geometry) requires a recalibrated value of $\langle f\rangle$ that
takes into account the UV--optical lag.

Single-epoch SMBH mass estimates are also unaffected by this result,
since the radius-luminosity (RL) relation is inferred from a sample of
RM AGN.  While the larger BLR radius measured from the UV data would
increase the normalization of the RL relation, a recalibration of
$\langle f\rangle$ exactly cancels this change.  The UV--optical lag
may introduce a second-order effect on single-epoch SMBH masses, if it
is found that the magnitude of the UV--optical lag correlates with
continuum luminosity or SMBH mass.  Furthermore, the magnitude of the
lag depends on accretion rate (see \S5.3), which may also add scatter
to existing mass-scaling relations.  To investigate these effects,
more simultaneous UV and optical RM experiments must be executed,
using a sample of AGN with a wide range of luminosities.

Finally, the UV--optical lag has an impact on masses derived from
direct dynamical modeling of RM data, since this method interprets the
continuum-line lag as a measure of the physical radius of the BLR.  To
a low approximation, a larger BLR radius implies a proportionally
larger SMBH mass.  The effect of using UV continuum light curves for
dynamical modeling studies will be investigated in future papers in
this series, but until such modeling is complete, we adopt a RM-based
SMBH mass for NGC 5548, since this estimate is less model dependent.
From the H$\beta$ virial products compiled by \citet{Bentz2015}, and
taking $\langle f\rangle = 4.3 \pm 1.1$ \citep{Grier2013}, we adopt a
mass of $(5.2\pm 1.3)\times 10^{7}\, {\rm M}_{\odot}$ for the SMBH in
NGC 5548.  We note that this value moves in the correct direction for
a larger BLR, but is still consistent within the quoted uncertainties
of the dynamically modeled mass in \citet{Pancoast2014}.

\subsubsection{Optical Smoothing}
The second difference between the UV and optical continuum light
curves is that the \emph{V}-band light curve appears to be smoother
than the \emph{HST} light curve.  For example, the rapid oscillations
in the UV light curve between HJD = 2,456,760 and 2,456,810 also
appear in the \emph{V}-band light curve, but at a much smaller
amplitude with gentler inflections.  The smoothing becomes
increasingly severe at longer wavelengths where the amplitude of
short-timescale variations decreases (see \S2.4).  These effects were
also seen in NGC 2617 by \citet{Shappee2014}, NGC 6814 by
\citet{Troyer2015}, and MCG-6-30-15 by \citet{Lira2015}.  Increased
smoothing and decreased amplitudes are expected if shorter-wavelength
emission drives the optical continuum, since the size, structure, and
inclination of the accretion disk define a ``continuum transfer
function'' that smooths the reprocessed light curve, while geometric
dilution decreases the energy flux incident on large disk radii that
contribute most to longer-wavelength emission.

In practical terms, the sharpest and strongest features in the
\emph{V}-band AGN STORM light curve are only slightly affected by this
smoothing. Since these features provide the most leverage for
constraining the CCF \citep{Peterson1993}, we conclude that the
smoothing of the optical continuum is not important for ground-based
RM studies that aim only to recover a mean emission-line lag and a
SMBH mass.  The smoothing may be more problematic for reconstructing
velocity-delay maps, direct dynamical modeling, or regularized linear
inversion \citep{Horne1991,Horne2004,Bentz2010,
  Grier2013,Pancoast2014,Skielboe2015}. These methods are very
sensitive to the fine structure of the driving continuum light curve,
and smoothing the light curve will erase information that would
otherwise be helpful for reconstruction of the geometry and dynamics
of the BLR.  Velocity-delay maps, dynamical modeling, and regularized
linear inversion for this dataset will be presented in upcoming
papers in this series.

\subsubsection{Magnitude of UV--Optical Lags}
The large lags measured for optical bands, shown in Figure
\ref{fig:lag_wv}, are comparable to, and sometimes larger than, the
lags for high-ionization-state lines such as He\,{\sc ii}\,$\lambda
1640$ and C\,{\sc iv}\,$\lambda 1549$ (Paper I).  If the lags do in
fact represent light-travel times across the accretion disk, then the
optically emitting portion of the accretion disk appears to have a
similar physical extent as the highly ionized portion of the BLR.
This situation implies a close connection between the BLR and
continuum-emitting source.  For example, BLR clouds may be directly
above or interior to the portion of the accretion disk emitting in the
optical.  Another plausible hypothesis is that at least part of the
inner, high-ionization BLR emission arises from a wind launched from
the surface of the accretion disk \citep[e.g.,
][]{Collin-Souffrin1987, Chiang1996,Proga2010}.  Such models are able
to reasonably explain many observed features of AGN emission lines,
including their profiles, variability, and absorption characteristics
\citep[see ][and references therein]{Proga2004, Eracleous2009,
  Denney2012, Higginbottom2014}.  Alternatively, the accretion disk
may smoothly merge with the BLR somewhere near 2--3 light days (for an
analysis of this family of models, see, for example,
\citealt{Goad2012}). Future papers in this series will attempt to map
the geometry and kinematics of the inner BLR using the reverberation
signal of high-ionization-state lines, which may shed further light on
the connection between the accretion disk and BLR.

\subsection{BLR Emission and Broad-Band Filter Lags}

Based on our spectral decomposition, approximately 19\% of the
observed emission in the \emph{u {\rm and} U} bands is Balmer
continuum emission from the BLR, while 15--20\% of \emph{r {\rm and}
  R}-band emission is the prominent H$\alpha$ line.  These
ratios may change with time, as shown in Table \ref{tab:fracs},
depending on the luminosity state of the AGN, the difference in phase
between the continuum and line light curves, and the light curves'
variability amplitudes.  For mean flux levels near the BLR
contamination in the \emph{u, U, r, {\rm and }R} bands, as well as
variability amplitudes and line lags that match the observed H$\beta$
light curve, our experiments with mock light curves indicate biases in
the interband continuum lag of $\sim\! 0.6$--$1.2$ days.

These results depend on the assumption that all BLR emission light
curves have properties similar to the H$\beta$ light curve.  It is
likely that the diffuse continuum actually has a stronger response but
smaller lag than H$\beta$, while H$\alpha$ is expected to have a
weaker response but larger lag
\citep{Korista2001,Korista2004,Bentz2010b}.  Since these parameters
have offsetting effects, it is unlikely that the lag biases caused by
BLR contamination are larger than the fiducial estimates presented
here (see Figure \ref{fig:sim_lag1}).  Future RM programs can test
this result by specifically targeting the diffuse continuum and
H$\alpha$ emission, putting stronger constraints on their variability
amplitudes and mean lags.

The systematic tendency for the \emph{u, U, r, {\rm and }R} band lags
to sit above the fit in Figure \ref{fig:lag_wv} can therefore
reasonably be explained by BLR contamination.  In the case of the {\it
  u} and {\it U} bands, the offset from the fit to Equation
\ref{equ:lag_wv} is large compared to the predicted lag (as well as
the observational uncertainty), which supports our decision to exclude
these data from the final model.  On the other hand, the {\it r} and
{\it R}-band offsets are much smaller, so the BLR bias probably makes
little difference for our final model.  Extending this reasoning to
the \emph{B, g, {\rm and} V}-band filters, the BLR contamination is
less than 10\%, which would result in even smaller biases.

It is therefore unlikely that there are any important biases of the
continuum lags in these bands, unless the diffuse continuum component
(e.g., free-free emission or the Paschen continuum) makes a
substantial contribution.  This diffuse continuum component of the
spectrum is unconstrained in our spectral decomposition, but it 
provides an intriguing possibility of explaining the downturn of the
lag-wavelength relation in the {\it I} and {\it z} bands.  The Paschen
continuum begins at 8204\,\AA, between the {\it i} and {\it I} bands,
so the true continuum lag-wavelength relation may run through the UV
and {\it Iz}-band lags, but underneath the lags of the other
optical filters.  The viability of this explanation requires
significant contamination of the optical filters by diffuse BLR
emission, which can potentially be estimated through photoionization
modeling of the {\it HST} data or additional optical/near-IR
observations.

\subsection{Accretion-Disk Size}
A geometrically thin, optically thick, irradiated accretion disk makes
definite predictions about the observed lag-wavelength structure of
the AGN.  Here, we compare this model to the observed continuum lags,
although we do not necessarily interpret the model parameters as
indicative of physical conditions within the AGN.  Full physical
modeling of the AGN STORM data is deferred to future papers in this
series (Starkey et al., in prep.; Kochanek et al., in prep.).

The disk is assumed to have a fixed aspect ratio with scale height
much smaller than radius, and is heated internally by viscous
dissipation and externally by a UV/X-ray source near the SMBH at a
small height $H$ above the disk.  In such a scenario, the temperature
profile is
\begin{align}
\label{Tprofile}
T(R) = \left(\frac{3GM\dot M}{8\pi \sigma R^3} + \frac{(1-A)L_{\rm X} H }{4\pi
    \sigma R^3} \right)^{1/4},
\end{align}
where $M$ is the mass of the central SMBH, $\dot M$ is the mass
accretion rate of the disk, $R$ is the distance away from the black
hole and central source of heating radiation, $L_{\rm X}$ is the
luminosity of the heating radiation, and $A$ is the albedo of the disk
\citep{Cackett2007}.  Here, we have ignored the inclination and the
inner edge of the disk, as well as any relativistic effects.
Inclination and relativity may have a small impact on the temperature
profile, but the largest effect is caused by the inner edge, which
reaches a maximum temperature and probably makes important
contributions to emission at wavelengths $<2000$\,\AA\
\citep{Novikov1973}.  This introduces an error when comparing the {\it
  HST} lags to this model, although the effect is small relative to
the UV--optical lags.

Identifying the temperature with a characteristic emission wavelength
$T= Xhc/k\lambda$, where $X$ is a multiplicative factor of order
unity, and the radius with the light-travel time $R = c\tau$, we have
\begin{align}
  c\tau =
  \left(X\frac{k\lambda}{hc}\right)^{4/3}\left(\frac{3GM\dot
      M}{8\pi \sigma } + \frac{(1-A)L_{\rm X} H }{4\pi \sigma} \right)^{1/3}.
\end{align}
The factor $X$ accounts for systematic issues in the conversion of $T$
to $\lambda$ for a given $R$, since a range of radii contributes to
emission at $\lambda$.  From the
flux-weighted mean radius
\begin{align}
\langle R \rangle = \frac{\int_{R_0}^{\infty} B(T(R))R^2\,dR}{\int_{R_0}^{\infty} B(T(R))R\,dR},
\end{align}
we derive $X = 2.49$, where $R_0$ is the inner edge of the disk,
$B(T(R))$ is the Planck function, and $T(R)$ is the temperature
profile defined in Equation \ref{Tprofile}.\footnote{Alternative
  definitions of $R$ exist.  For example, a weighting function that
  better characterizes the radius responding to variable irradiation
  would replace Equation \ref{Tprofile} with $T = T_0(R) +
  \frac{\partial B(T(R))}{\partial T}\frac{\delta T}{T}T$, and set
  $\frac{\delta T}{T}$ equal to a constant fractional temperature
  variation.  This yields $X = 3.37$.}

If we measure $\tau$ relative to a reference time delay $\tau_0$ of a
light curve with effective wavelength $\lambda_0$, then this becomes
\begin{align}
\label{equ:MmdotLx}
(\tau - \tau_0) =
\frac{1}{c}\left(X\frac{k\lambda_0}{hc}\right)^{4/3}\left(\frac{3GM\dot M}{8\pi
    \sigma } + \frac{(1-A)L_{\rm X} H }{4\pi \sigma}  \right)^{1/3}\nonumber \\
\left[\left( \frac{\lambda}{\lambda_0} \right)^{4/3} - 1  \right].
\end{align}
Therefore, the parameter $\alpha$ in Equation \ref{equ:lag_wv} is
related to the energy generation rate responsible for heating the
disk, while $\beta$ is predicted to be $4/3$.  The absolute size of
the disk at $\lambda_0$ can be measured by determining $\tau_0$, which
is inferred by assuming the corona is located at $\tau = 0$ and
fitting the X-rays lags (in which case $\tau_0 = \alpha$).

We can only determine $\dot M$ indirectly through an estimate of the
bolometric luminosity.  We set $L_{\rm Bol} = \eta\dot Mc^2$, where
$\eta$ is the radiative efficiency for converting rest mass into
radiation, and $L_{\rm Bol}$ quantifies all emergent radiation from
the AGN, including coronal X-rays (in this sense, our model differs
from the typical \citealt{Shakura1973} thin-disk model).  A convenient
parameterization of $L_{\rm Bol}$ is the Eddington ratio, $\dot m_{\rm
  E} = L_{\rm Bol}/L_{\rm Edd} $.  We also simplify Equation
\ref{equ:MmdotLx} by taking $(1-A)L_{\rm X} H/R = \kappa G M \dot
M/2R$, where $\kappa$ is the local ratio of external to internal
heating, assumed to be constant with radius.

The equation for $\alpha$ is then
\begin{align}
\alpha = \frac{1}{c}\left(X\frac{k\lambda_0}{hc} \right)^{4/3}\left[\left( \frac{GM}{8\pi \sigma}\right)\left(\frac{L_{\rm Edd}}{\eta c^2}\right)\left(3 + \kappa\right)\dot m_{\rm E}\right]^{1/3}.\label{equ:alpha}
\end{align}
A common choice for $\dot m_{\rm E}$ is 0.1, and we further assume
that $\eta = 0.1$ and $\kappa = 1$ for our fiducial calculations
(i.e., the X-rays and viscous heating contribute equal amounts of
energy to the disk).  For a $5.2\times 10^{7}\text{ M}_\odot$ SMBH,
these assumptions give $\alpha = 0.14$ day.  If we increase the
accretion rate by setting $\dot m_{\rm E} = 1$ and 10, then $\alpha =
0.30$ and 0.65 day, respectively.  The lag-wavelength relation for these
models is shown in Figure \ref{fig:lag_wv}, and the curves for $\dot
m_{\rm E} = 1$--10 bracket our fit to Equation \ref{equ:lag_wv} with
$\alpha = $\ \alphafit\ days and $\beta = $\ \betafit.  However, it is
important to note that the disk probably does not remain geometrically
thin at these high accretion rates, and the assumptions of the model
do not hold in this regime \citep{Jiang2014,Sadowski2014}.  Equation
\ref{equ:alpha} is relatively insensitive to the ratio of external to
internal heating---even if the X-rays contribute a negligible portion
of the luminosity ($\kappa = 0$), $\alpha$ would only change by a
factor of $(3/4)^{1/3}$.

\begin{figure}
  \includegraphics[width=0.5\textwidth]{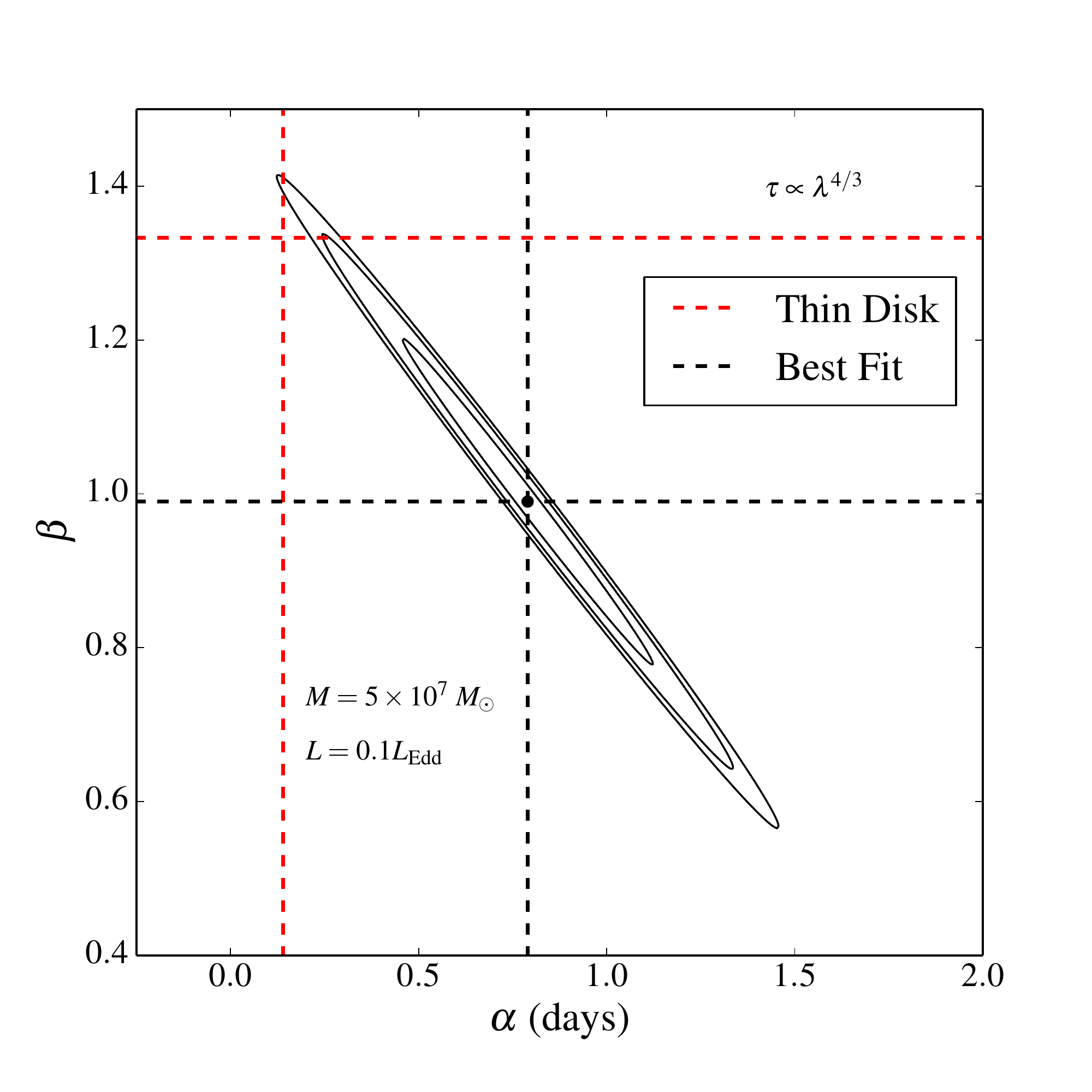}
  \caption{Probability distribution for the parameters $\alpha$ and
    $\beta$ of our best-fit model.  The contours show the 68, 95, and
    99\% confidence regions.\label{ref:corparam}}
\end{figure}

In Paper II, we found $\alpha = 0.35\pm 0.04$ day, somewhat smaller
than in this study.  The smaller value can be explained by
correlations between $\alpha$ and $\beta$, shown in Figure
\ref{ref:corparam}.  For the final analysis in Paper II, $\beta$ was
fixed to 4/3, and, if we do the same, we find $\alpha = 0.42 \pm 0.02$
day, in good agreement with Paper II.  The fit with fixed $\beta$ has
$\chi^2/{\rm dof} = 1.42$, making the lower value of $\beta$=\
\betafit\ statistically preferred.  However, this result is driven by
the flattening of the lags at the reddest wavelengths.  If we exclude
the \emph{I \emph{\rm and} z} bands from the fit (as well as \emph{u
  {\rm and} U}), we find $\beta = 1.18 \pm 0.19$ and $\alpha =
0.58\pm 0.20$\ day with $\chi^2/{\rm dof} = 0.89$, while fixing
$\beta = 4/3$ gives $\alpha = 0.45\pm0.02$\ day and $\chi^2/{\rm dof}
= 0.87$.

We therefore conclude that a reprocessing model can fit the data
reasonably well but requires a much larger disk radius than predicted
by standard thin-disk models.  Fixing $\beta = 4/3$ (in order to match
the theoretical temperature profile), our best-fit value of $\alpha =
0.42$ day is a factor of 3.0 larger than the standard prediction with
$L/L_{\rm Edd} = 0.1$.

A sufficiently high accretion rate can account for this difference by
increasing the size of the accretion disk.  We note that uncertainties
in the SMBH mass do not require $\dot m_{\rm E}$ to be larger than
one, since $\dot m_{\rm E} \propto \alpha^3/M^2$ (Equation
\ref{equ:alpha}), while the SMBH mass may be up to 1.75 times larger
at 3$\sigma$ than our adopted value.  This would still require $\dot
m_{\rm E}$ to be somewhere in the range $\sim\! 0.1$--1.  On the other
hand, a comparison of $\dot m_{\rm E}$ can be made assuming a 
thin-disk spectrum and using the observed optical luminosity
\citep[e.g.,][Equation 4.53]{Collin2002, Netzer2013}.  From our
spectral decompositions, we estimate that $\lambda F_{\lambda} =
4.57\times 10^{-11}$ erg cm$^{-2}$ s$^{-1}$ at 5100\,\AA, which yields
$\dot m_{\rm E} = 0.05 $ at a disk inclination of $\cos i = 0.63$ and
radiative efficiency $\eta = 0.1$.  The accretion rate cannot be much
higher (unless the disk is seen very edge-on), so this result implies
that a standard thin-disk model cannot account for both the observed
time delays and the monochromatic luminosity at 5100\,\AA.

The large disk size found here corroborates the results from Paper II
and other recent RM studies \citep{Shappee2014,McHardy2014,Lira2015}.
The measurements of large disk radii are also in good agreement with
the sizes inferred from gravitational microlensing experiments (see
Figure 6 in Paper II, as well as \citealt{Poindexter2008, Morgan2010,
  Mosquera2013}).  Other sources of tension with the
thin-disk/continuum reprocessing model are (1) the weak correlation
between the X-ray light curves and UV/optical light curves (Paper II),
and (2) the possible flattening of the lags at the longest wavelengths.
The latter phenomenon might contain information about the outer edge
of the disk, or perhaps be explained by contaminating emission from
BLR material along the line of sight \citep{Korista2001}, and/or
emission from the inner edge of the near side of the obscuring torus
\citep{Goad2012}.

The intriguing result that accretion disks in AGN might be larger than
predicted by standard thin-disk theory depends on only a handful of
lensed quasars and three RM AGN (NGC 5548, NGC 2617, and MCG-6-30-15;
\citealt{Shappee2014,Lira2015}).  Thus, it is important to carry out
further continuum RM experiments, in order to establish if this is a
robust result and determine what physical parameters govern the disk
size.  It is also possible to recast this kind of experiment in more
direct scaling relations, such as the lag-luminosity relations of
\citet{Sergeev2005}, which can be derived from thin-disk theory (both
the disk size and luminosity scale with accretion rate and black hole
mass).  In fact, the \citet{Sergeev2005} lag-luminosity relations lie
somewhat above the prediction for standard thin-disk theory, and the
lags reported here would be $\sim$1 day below these relations in most
bands.  However, the relations are largely based on unresolved lags
and have very large uncertainties, so they do not put an interesting
constraint on model predictions.  A larger sample of AGN with
continuum lags derived to the same precision as this study would
provide an interesting measurement of the lag-luminosity relations,
which can provide a further test of thin-disk theory and establish if
larger disk sizes are generic properties of the AGN population.

\section{Summary}
We have presented results for a ground-based, broad-band photometric
monitoring campaign of NGC 5548.  Our light curves are of very high
quality, achieving cadences of $\lesssim$ 1 day in nine optical bands
over an entire observing season.  Using full optical-wavelength
spectra and synthetic photometry, we estimated the relative
contribution of host-galaxy starlight, AGN continuum emission, Balmer
continuum, and line emission from the BLR to the observed light
curves.  Our main results are as follows.

\begin{enumerate}[1]
\item Significant time delays are detected between the far-UV, near-UV, 
  and optical broad-band light curves.  The delay between emission
  at \drivelc\ and 2600\,\AA\ is less than 1 day, and the delay
  between emission at \drivelc\ and the {\it V} band is about 2
  days. Such large time delays are comparable to, and sometimes
  greater than, the lags of the high-ionization-state emission lines,
  suggesting that the continuum-emitting source is of a physical size
  approximately equal to the inner BLR.

\item If similar interband continuum lags exist in other AGN, this also
  suggests that the size of the BLR is 11--37\% larger than would be
  inferred from optical data alone.  However, there do not appear to
  be other significant systematic effects associated with the optical
  light curves, and RM SMBH masses are not affected by this result.

\item There is some contamination of the broad-band light curves by
  BLR emission, with $19$\% of the \emph{u {\rm and} U} bands
  attributable to the Balmer continuum, and 15--20\% of the \emph{r
    {\rm and} R} bands attributable to H$\alpha$.  The impact of BLR
  emission on the observed {\it u} and {\it U}-band lags is $\sim\!
  0.6$--$1.2$ days, but is probably unimportant in the {\it r} and
  {\it R} bands. This justifies our decision to exclude the {\it u}
  and {\it U}-band lags from our final analysis.  

\item The trend of lag with wavelength is broadly consistent with the
  prediction for continuum reprocessing by a geometrically thin
  accretion disk with $\tau \propto \lambda^{4/3}$.  However, the size
  of the disk is a factor of 3 larger than the prediction for standard
  thin-disk theory, assuming that $L = 0.1\, L_{\rm Edd}$.  This result
  appears to corroborate those from other continuum RM projects and
  gravitational microlensing studies.  Further investigations of the
  accretion-disk structure will benefit from physical modeling of the
  AGN STORM light curves, and several such studies are planned for
  upcoming papers in this series (Starkey et al., in prep.; Kochanek et
  al., in prep.).

\end{enumerate}

The LBT is an international collaboration among institutions in the
United States, Italy and Germany. LBT Corporation partners are: The
Ohio State University, and The Research Corporation, on behalf of The
University of Notre Dame, University of Minnesota and University of
Virginia; The University of Arizona on behalf of the Arizona
university system; Istituto Nazionale di Astrofisica, Italy; LBT
Beteiligungsgesellschaft, Germany, representing the Max-Planck
Society, the Astrophysical Institute Potsdam, and Heidelberg
University.

This paper used data obtained with the MODS spectrographs built with
funding from National Science Foundation (NSF) grant AST-9987045 and
the NSF Telescope System Instrumentation Program (TSIP), with
additional funds from the Ohio Board of Regents and the Ohio State
University Office of Research.  This paper made use of the {\tt
  modsIDL} spectral data reduction pipeline developed in part with
funds provided by NSF Grant AST - 1108693.

The Liverpool Telescope is operated on the island of La Palma by
Liverpool John Moores University in the Spanish Observatorio del Roque
de los Muchachos of the Instituto de Astrofisica de Canarias with
financial support from the UK Science and Technology Facilities
Council.

KAIT and its ongoing operation were made possible by donations from
Sun Microsystems, Inc., the Hewlett-Packard Company, AutoScope
Corporation, Lick Observatory, the NSF, the University of California,
the Sylvia and Jim Katzman Foundation, and the TABASGO Foundation.
Research at Lick Observatory is partially supported by a generous gift
from Google.

Support for \HST\ program number GO-13330 was provided by NASA through
a grant from the Space Telescope Science Institute,
which is operated by the Association of Universities for Research in
Astronomy, Inc., under NASA contract NAS5-26555.
M.M.F., G.D.R., B.M.P., C.J.G., and R.W.P.\ are grateful for the support of the
NSF through grant AST-1008882 to The Ohio
State University. 
A.J.B.\ and L.P.\ have been supported by NSF grant AST-1412693.
A.V.F.\ and W.-K.Z.\ are grateful for financial assistance from NSF grant
AST-1211916, the TABASGO Foundation, and the Christopher R. Redlich Fund. 
M.C.\ Bentz gratefully acknowledges support through NSF CAREER grant AST-1253702 to Georgia State University.
M.C.\ Bottorff acknowledges HHMI for support through an undergraduate science education grant to
Southwestern University.
K.D.D.\ is supported by an NSF Fellowship awarded under grant AST-1302093.
R.E.\ gratefully acknowledges support from NASA under awards NNX13AC26G, NNX13AC63G, and NNX13AE99G.
J.M.G.\ gratefully acknowledges support from NASA under award NNH13CH61C.
P.B.H.\ is supported by NSERC. 
M.I.\ acknowledges support from the Creative Initiative program, No. 2008-0060544, of the National Research Foundation of Korea (NRFK) funded by the Korean government (MSIP).
M.D.J.\ acknowledges NSF grant AST-0618209 used for obtaining the 0.91 m telescope at WMO.  
SRON is financially supported by NWO, the Netherlands Organization for Scientific Research.
B.C.K.\ is partially supported by the UC Center for Galaxy Evolution.
C.S.K.\ acknowledges the support of NSF grant AST-1009756.  
D.C.L.\ acknowledges support from NSF grants AST-1009571 and AST-1210311,
under which part of this research (photometric observations collected at
MLO) was carried out.
We thank Nhieu Duong, Harish Khandrika, Richard
Mellinger, J. Chuck Horst, Steven Armen, and Eddie Garcia for assistance
with the MLO observations.
P.L.\ acknowledges support from Fondecyt grant \#1120328.
A.P.\ acknowledges support from a NSF graduate fellowship, a UCSB Dean's
Fellowship, and a NASA Einstein Fellowship.
J.S.S.\ acknowledges CNPq, National Council for Scientific and Technological Development (Brazil)
for partial support and The Ohio State University for warm hospitality.
T.T.\ has been supported by NSF grant AST-1412315.
T.T.\ and B.C.K.\ acknowledge support from the Packard Foundation in the form of a Packard Research
Fellowship to T.T.; also, T.T. thanks the American Academy in Rome and the Observatory
of Monteporzio Catone for kind hospitality.
The Dark Cosmology Centre is funded by the Danish National Research Foundation.
M.V.\ gratefully acknowledges support from the Danish Council for Independent Research via grant no. DFF--4002-00275.
J.-H.W.\ acknowledges support by the National Research Foundation of Korea (NRF) grant funded by the Korean government (No. 2010-0027910).
E.D.B.\ is supported by Padua University through grants 60A02-5857/13, 60A02-5833/14, 60A02-4434/15, and
CPDA133894.
K.H. acknowledges support from STFC grant ST/M001296/1.
S.A.K. thanks Dr. I. A. Rakhimov, the Director of Svetloe Observatory, for his support and hospitality.

This research has made use of
the NASA/IPAC Extragalactic Database (NED), which is operated by the
Jet Propulsion Laboratory, California Institute of Technology, under
contract with NASA.

\bibliography{ms_change.bbl}

\section*{Appendix---On Interpolation}  

Interpolating a light curve requires some assumed model, which may be more or less
sophisticated.  For example, linear interpolation is a very simple
method.  However, linear interpolation assumes no additional
variability between sampled epochs, and this is known to be an
incorrect description of AGN light curves on nightly timescales.  The
DRW allows for intrinsic variations between sampled epochs by modeling
the data covariance, from which we can make a better guess as to what
the continuum is doing between the observations and, moreover, assign
a meaningful error bar to the prediction.

\begin{figure}
\includegraphics[width=0.5\textwidth]{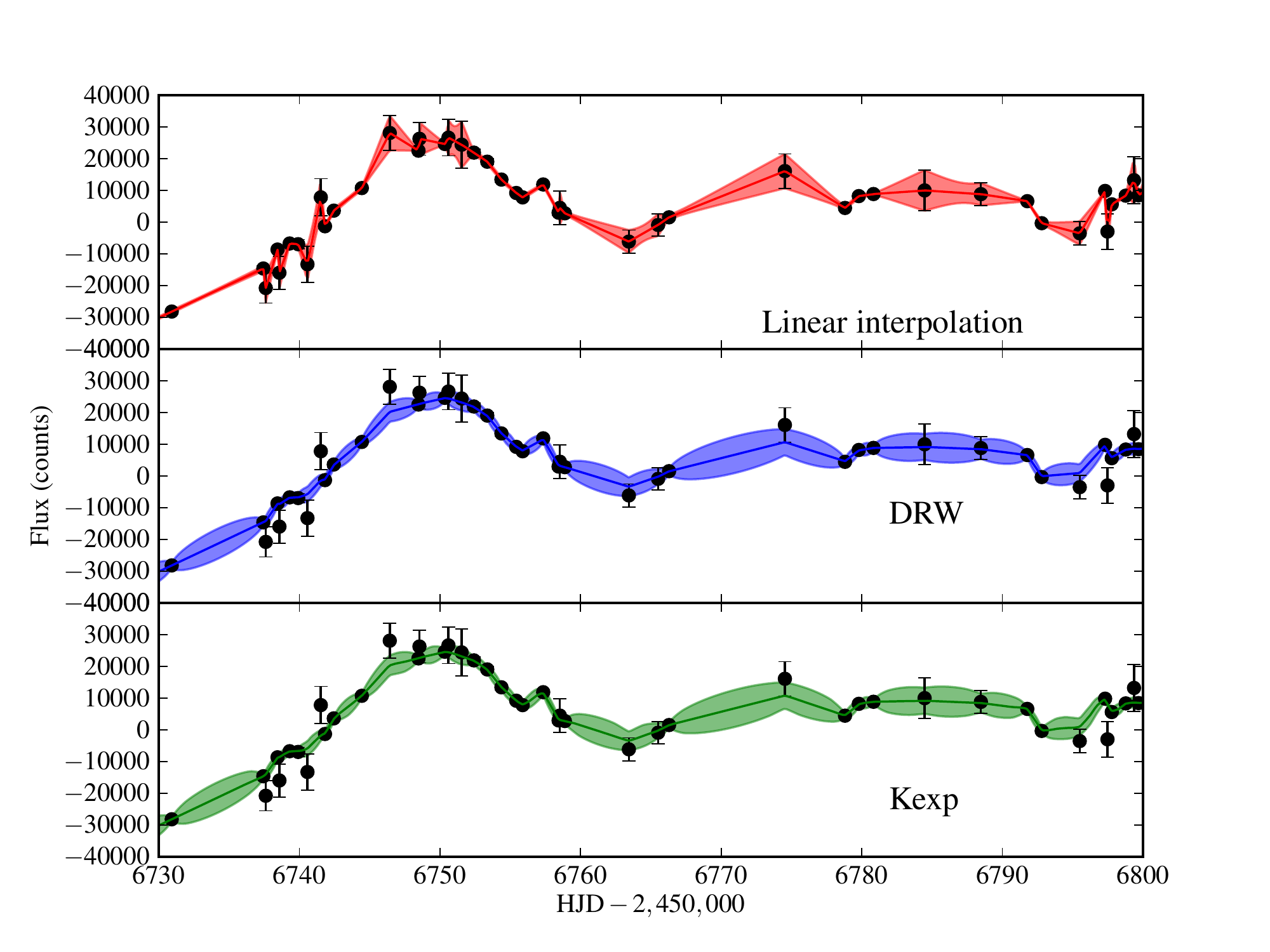}
\caption{Data and models for the $R$-band continuum in NGC
  5548. Top panel: Linear interpolation model. Middle panel: DRW
  model. Bottom panel: ``Kepler-exponential model,'' which is DRW with
  a drop in power on the shortest timescales ($\tau_{\rm cut} = 1$\ day).  \label{interp_plot} }
\end{figure}

  Figure \ref{interp_plot} shows the
  linear interpolation model of a portion of the $R$-band continuum
  light curve of NGC 5548 (the {\it R} band was chosen because it has
  large gaps). By definition, the linearly interpolated model goes
  exactly through every data point. It has an ``error snake'' that
  matches the error bars of the data at a sampled epoch and can shrink
  between data points. This is because the model includes only
  measurement noise, so the error in the model is smallest somewhere
  in between the data points where it best averages the two
  measurements.  Defining the fractional distance between the
  interpolated epoch $t_{j}$ and the data points $t_i$ and $t_{i+1}$
  as $x = (t_j - t_i)/(t_{i+1 } -t_i)$ so $0 < x < 1$, the error snake
  for linear interpolation at $t_j$ is given by $\sigma^2 (t_j) = (1 -
  x)^2\sigma (t_i)^2 + x^2\sigma (t_{i+1})^2$, which is smallest at $x =
  1/2$ for $\sigma (t_i) =\sigma(t_{i+1})$. Because it is required to
  go through the data points, there are regions of the model light
  curve (e.g., near day 6740) where the model rapidly ``oscillates'' in
  order to pass through nearby points.  The principal problems with
  the linear interpolation model are therefore (1) that the model
  light curve has much more structure than it should when the light
  curve is well sampled, and (2) the error snake can decrease in width
  the farther it gets from the actual data points.  

  {\tt JAVELIN} uses a covariance model to estimate the statistical
  properties of light curves.  We have used the DRW model because it
  is simple and describes quasar variability on the timescales sampled
  by the data \citep{Kelly2009,MacLeod2010,Zu2013}.  The middle panel
  of Figure \ref{interp_plot} shows the DRW model for the same region
  of the NGC 5548 light curve. There are two important qualitative
  changes. First, unlike linear interpolation, the model no longer has
  to go through the data points. For example, in the region near day
  6740, the DRW model is quite smooth because it has decided
  (statistically) that the three points with larger uncertainties
  should be viewed as measurement fluctuations rather than intrinsic
  variability. In contrast, there is the region near day 6758 where
  the error bars on the points forming a ``triangle'' are small enough
  that the model tracks the data points more or less like the linear
  interpolated model. The second difference is that the error snakes
  generally grow in the gaps between the data points.  This is because
  {\tt JAVELIN} is accounting for the intrinsic variability as well as
  the measurement errors. The more distant an actual measurement, the
  greater the expected variance in the underlying light curve. If the
  measurement errors are very large (e.g., the point near day 6785),
  then the error snake can be smaller than the measurement errors
  because the model predicts the expected range of the light curve
  better than it was actually measured.  

 There is evidence from high-cadence {\it Kepler} light curves
  that the DRW model overestimates the variability power on short
  (subweek) timescales \citep{Edelson2014,Kasliwal2015}.  Although the
  DRW is therefore an incorrect model on short timescales, our data
  have very different properties from the {\it Kepler} light curves (1
  day cadence instead of 30 minute cadence and $\sim \! 0.5$--$1.0$\%
  uncertainties instead of $\sim\!  0.1$\%), and useful results can be
  found as long as the covariance model is a reasonable approximation
  of the true data covariance.  An analogy exists here with optimal
  (Weiner) filters---quoting from {\it Numerical Recipes}
  (\citealt{Press2002}, Chapter 13.3, page 651), ``In other words,
  even a fairly crudely determined optimal filter can give excellent
  results when applied to data.''  This is because errors in the
  covariance model only become significant when the differences are
  larger than the noise $\sigma$.  For two structure-function
  amplitudes $SF_1$ and $SF_2$, the fractional changes in the models
  are of order $| SF_1^2 - SF_2^2 |/\sigma^2$.  Unless the light curve
  is of sufficiently high quality to measure the structure function on
  a given timescale, we will not have any noticeable effects from
  making even order unity errors in the structure function on those
  timescales.  To go back to the optimal filtering analogy, we get
  90\% of the gains from being in the ball park, and very little extra
  from being perfectly correct.

We can illustrate this by using the ``Kepler-exponential'' model from
\citet{Zu2013}, which includes a timescale $\tau_{\rm cut}$ below
which the power spectrum is cut off.  The Kepler-exponential model was
designed to explore the {\it Kepler} results (that AGN light curves
have suppressed power on short timescales), and is available as an
option in {\tt JAVELIN} (the {\tt JAVELIN} algorithm can use any
covariance model desired).  The Kepler-exponential model of the {\it
  R}-band light curve is shown in Figure \ref{interp_plot} with a
power cutoff timescale $\tau_{\rm cut} = 1$\ day.  As expected, it is
very difficult to see any differences. The easiest one to spot is that
the error snake grows a little faster as it moves away from a data
point in the DRW model because it has some extra small-scale power
(which actually makes the DRW model a more conservative choice).  If
we steadily increase $\tau_{\rm cut}$ above 1 day, the
Kepler-exponential models start to fit the data poorly because they
have too little short-timescale power.

We compared the DRW and Kepler-exponential models quantitatively by
assessing how well they predict the data.  We generated predicted
values for each data point from the interpolation scheme described in
\S2.3 for each model, and then calculated
\begin{align*}
\chi^2/{\rm dof} = \frac{1}{N-k}\sum_i^N \frac{(y_i - m_i)^2}{\sigma_i^2},
\end{align*}
where $N$ is the number of data points, $k$ is the number of
parameters, $y$ are the data, $m$ are the interpolated values, and
$\sigma$ are the uncertainties (on the data only---the uncertainty on
the interpolation is necessarily consistent with the data).  We
use all data points when calculating the interpolation, and so we emphasize 
that this definition has nothing to do with the probability
of the model: linear interpolation would force this value of
$\chi^2/{\rm dof}$ = 0, even though it is certainly not correct.
Rather, this definition gives an estimate of the consistency of the
data with the model.  For these fits, we again fixed $\tau_{\rm cut}$
to 1 day.

Table \ref{Kexp_comp} summarizes these results.  The two models
produce interpolations that are virtually indistinguishable (i.e.,
nearly equal $\chi^2/{\rm dof}$).  Increasing $\tau_{\rm cut}$ to 10
days increases $\chi^2/{\rm dof}$ by a small amount (up to 0.06), and
as $\tau_{\rm cut}$ approaches 0 days, we recover the DRW.  This means
that there is no quantitative advantage to using a random process with
suppressed short-timescale power---our data are not good enough to see
this effect.  

\begin{centering}
\begin{deluxetable}{c|cc}
\tablewidth{0pt}
\tablecaption{Comparison of DRW and Kepler-Exponential Interpolations.\label{Kexp_comp}}
\tablehead{ 
\colhead{Band} & \colhead{$\chi^2/{\rm dof}$\ (DRW)} & \colhead{$\chi^2/{\rm dof}$\ (Kexp)}}
\startdata
{\it u} & 0.37 & 0.34  \\
{\it B} & 0.29 & 0.29  \\
{\it g} & 0.33 & 0.31  \\
{\it V} & 0.65 & 0.73  \\
{\it r} & 0.40 & 0.40  \\
{\it R} & 0.41 & 0.41  \\
{\it i} & 0.23 & 0.22  \\
{\it I} & 0.28 & 0.26  \\
{\it z} & 0.27 & 0.26  \\
\enddata
\end{deluxetable}
\end{centering}

\end{document}